\documentclass[12pt]{article}
\usepackage{ametsoc}
\usepackage{amsmath} 
\usepackage{amssymb} 
\usepackage{amsfonts}
\usepackage{graphicx,color}                 
\usepackage{epsfig}
\usepackage{lineno}
\usepackage{subfigure}

\newcommand{\myabstract}{
  This study considers the data assimilation problem in coupled systems, which consists of two components (sub-systems) interacting with each other through certain coupling terms. A straightforward way to tackle the assimilation problem in such systems is to concatenate the states of the sub-systems into one augmented state vector, so that a standard ensemble Kalman filter (EnKF) can be directly applied. In this work we present a divided state-space estimation strategy, in which data assimilation is carried out with respect to each individual sub-system, involving quantities from the sub-system itself and correlated quantities from other coupled sub-systems. On top of the divided state-space estimation strategy, we also consider the possibility to run the sub-systems separately. Combining these two ideas, a few variants of the EnKF are derived. The introduction of these variants is mainly inspired by the current status and challenges in coupled data assimilation problems, and thus might be of interest from a practical point of view. Numerical experiments with a multi-scale Lorentz 96 model are conducted to evaluate the performance of these variants against that of the conventional EnKF. In addition, specific for coupled data assimilation problems, two prototypes of extensions of the presented methods are also developed in order to achieve a trade-off between efficiency and accuracy.
}
\begin{document}
%

%
\title{\textbf{\large{Ensemble Kalman filtering with a divided state-space strategy for coupled data assimilation problems}}}
%
%
\author{\textsc{Xiaodong Luo}
				\thanks{E-mail: xiaodong.luo@iris.no}\\
				\textit{\footnotesize{International Research Institute of Stavanger (IRIS), 5008 Bergen, Norway}}\\
				\quad\textsc{and Ibrahim Hoteit}\thanks{\textit{Corresponding author.} E-mail: ibrahim.hoteit@kaust.edu.sa}\\
				\textit{\footnotesize{King Abdullah University of Science and Technology (KAUST), Thuwal, Saudi Arabia}}\\
}
%
\ifthenelse{\boolean{dc}}
{
\twocolumn[
\begin{@twocolumnfalse}
\amstitle

\begin{center}
\begin{minipage}{13.0cm}
\begin{abstract}
	\myabstract
	\newline
	\begin{center}
		\rule{38mm}{0.2mm}
	\end{center}
\end{abstract}
\end{minipage}
\end{center}
\end{@twocolumnfalse}
]
}
{
\amstitle
\begin{abstract}
\myabstract
\end{abstract}
}

\section{Introduction} \label{sec:introduction}

This work considers the data assimilation problem in coupled systems that consist of two sub-systems. Examples in this aspect include, for instance, coupled ocean-atmosphere models (e.g., \citealp{Russell1995}), marine ecosystem models coupling physics and biology (e.g., \citealp{petihakis2009eastern}), coupled flow and (contaminant) transport models (e.g., \citealp{Dawson2004-compatible}), to name a few.

In principle, data assimilation in coupled systems can be tackled by concatenating the states of the sub-systems into one augmented state and treating the whole coupled system as a single dynamical system. After augmentation, a conventional data assimilation method, such as the ensemble Kalman filter (EnKF), can be directly applied. 
In this work we present a divided state-space estimation strategy in the context of ensemble Kalman filtering. Instead of directly applying the update formulae in the conventional EnKF, we consider the possibility to express the update formulae in terms of some quantities with respect to the sub-systems themselves. In doing so, the update formulae in the divided estimation framework introduces some extra ``cross terms'' to account for the effect of coupling between the sub-systems. 

The divided estimation framework is derived based on the joint estimation one, hence in principle these two approaches are mathematically equivalent. The main purpose of this work is to investigate the possibility of using the divided estimation strategy as an alternative to its joint counterpart. Whenever convenient, we would advocate the use of the joint estimation strategy, since it is conceptually more straightforward. However, there might still be some aspects in which the divided estimation strategy may appear more attractive, e.g., in terms of flexibility of implementation in large-scale applications, as to be further discussed later.

This work is organized as follows. Section \ref{sec:ensemble_filtering} outlines the filtering step of the EnKF in the joint and divided estimation frameworks. Section \ref{sec:experiments} conducts numerical experiments with a multi-scale Lorenz 96 model, and verifies that the joint and divided estimation frameworks have close performance under the same conditions. Section \ref{sec:extensions} investigates two extensions of the divided estimation framework that aim to achieve a certain trade-off between computational efficiency and accuracy. 
Finally, Section \ref{sec:conclusion} concludes the work and discusses some potential future developments.

\section{Joint and divided estimation strategies with the EnKF} \label{sec:ensemble_filtering}
In the literature there are many variants of the EnKF, for example, see \citet{Anderson-ensemble,Bishop-adaptive,Evensen-sequential,Burgers-analysis,Hoteit2002,Luo-ensemble,Tippett-ensemble,Whitaker-ensemble}. In this work we use the ensemble transform Kalman filter (ETKF) \citep{Bishop-adaptive} for illustration. The extension to other filters can be done in a similar way. The joint and divided estimation strategies mainly differ at the filtering step, which is thus our focus hereafter. For ease of notation, we drop the time indices of all involved quantities. 

Suppose that the state vectors in the coupled sub-systems are $\mathbf{\eta}$ and $\mathbf{\xi}$, respectively, and the corresponding observation sub-systems are given by $\mathbf{y}_{\eta} = \mathcal{H}_{\eta} \, \mathbf{\eta} + \mathbf{u}_{\eta}$ and $\mathbf{y}_{\xi} = \mathcal{H}_{\xi} \, \mathbf{\xi} + \mathbf{u}_{\xi}$, where $\mathcal{H}_{\eta}$ and $\mathcal{H}_{\xi}$ are some linear observation operators \footnote{In cases of nonlinear observation operators, one may either approximate them by some linear ones, or adopt more sophisticated assimilation schemes (see, for example, \citealp{Hoteit2012,Luo2008-spgsf1,Luo2014ensemble,VanLeeuwen2009,Zupanski-maximum}).}, and $\mathbf{u}_{\eta}$ and $\mathbf{u}_{\xi}$ the corresponding observation noise with zero means and covariances $\mathbf{R}_{\eta}$ and $\mathbf{R}_{\xi}$, respectively. In practice it is possible that one of the sub-systems (e.g., $\mathbf{\xi}$) may not be observed. In this case, to overcome the technical problem in describing the unknown observation operator (e.g., $\mathcal{H}_{\xi}$), one can set the associated covariance matrix $\mathbf{R}_{\xi}$ of $\mathbf{y}_{\xi}$ to $+\infty$ so that $\mathbf{y}_{\xi}$ does not affect the update \citep[p.~219]{Jazwinski1970}. For convenience of discussion, we denote the dimensions of the vectors $\mathbf{\eta}$, $\mathbf{\xi}$, $\mathbf{x}$, $\mathbf{y}_{\eta}$, $\mathbf{y}_{\xi}$ and $\mathbf{y}$ by $m_{\eta}$, $m_{\xi}$, $m_{x}$, $m_{\eta}^{obv}$, $m_{\xi}^{obv}$ and $m_{y}$, respectively, such that $m_{\eta} + m_{\xi} = m_{x}$ and $m_{\eta}^{obv} + m_{\xi}^{obv} = m_{y}$.

In the above setting we have assumed that the observation operators $\mathcal{H}_{\eta}$ and $\mathcal{H}_{\xi}$ for different sub-systems are ``separable'', in the sense that the observation (say $\mathbf{y}_{\eta}$) of each sub-system only depends on the corresponding sub-system state (say $\eta$). In some situations, however, the observation with respect to one sub-system may depend on the state variables of both sub-systems. In such cases, one may introduce a certain transform to the observation system augmented by the observations with respect to the sub-systems (see Eq. (\ref{eq:augmented_observation_system})), so that the resulting augmented observation system (after the transform) has a diagonal or block diagonal observation operator, and thus becomes ``separable''.

One can concatenate the above observation sub-systems and obtain 
\begin{linenomath*}
\begin{equation} \label{eq:augmented_observation_system}
\mathbf{y} = \mathcal{H} \, \mathbf{x} + \mathbf{u} \, ,
\end{equation}
\end{linenomath*}
where $\mathcal{H} \, \mathbf{x} \equiv [(\mathcal{H}_{\eta} \, \mathbf{\eta} )^T,(\mathcal{H}_{\eta} \, \mathbf{\xi})^T]^T$, and $\mathbf{u} = [\mathbf{u}_{\eta}^T,\mathbf{u}_{\xi}^T]^T$ is the augmented observation noise with zero mean and covariance $\mathbf{R}$. Here
\begin{linenomath*}
\begin{equation} \label{eq:R_relation}
\mathbf{R} =
\begin{pmatrix}
\mathbf{R}_{\eta} & \mathbf{R}_{\eta \xi} \\
\mathbf{R}_{\eta \xi}^T & \mathbf{R}_{\xi} \\
\end{pmatrix} \, ,
\end{equation}
\end{linenomath*}
with $\mathbf{R}_{\eta \xi}$ being the cross-covariance between $\mathbf{u}_{\eta}$ and $\mathbf{u}_{\xi}$. Throughout this work, we assume the observation noise $\mathbf{u}_{\eta}$ and $\mathbf{u}_{\xi}$ are uncorrelated, such that $\mathbf{R}_{\eta \xi} = \mathbf{0}$. If, in addition, both $\mathbf{R}_{\eta}$ and $\mathbf{R}_{\xi}$ are diagonal, then the observation $\mathbf{y}$ can be assimilated serially through some scalar update formulae \citep{Anderson2003-local}. 
For our deduction, though, we only need to assume that $\mathbf{R}$ is a block diagonal matrix.
If this is not the case, i.e., $\mathbf{R}_{\eta \xi} \neq \mathbf{0}$, one can still obtain results similar to those presented below (though in somewhat more complicated forms), following a procedure similar to the derivation in Appendix \ref{sec:appendix A}.

Let $\mathbf{X}^b = \{\mathbf{x}_i^b: \mathbf{x}_i^b = [(\mathbf{\eta}_i^b)^T \, , (\mathbf{\xi}_i^b)^T ]^T \, , i = 1, \dotsb, n \}$ be an $n$-member background ensemble consisting of the sub-system components $\mathbf{\Phi}^b \equiv \{\mathbf{\eta}_i^b, i = 1, \dotsb, n \}$ and $\mathbf{\Xi}^b \equiv \{\mathbf{\xi}_i^b, i = 1, \dotsb, n \}$. In addition, let
\begin{linenomath*}
\begin{subequations}
\begin{align}
\label{eq:augmented_bg_mean} & \bar{\mathbf{x}}^b = \dfrac{1}{n} \sum\limits_{i=1}^{n} \mathbf{x}_i^b \, , \\
\label{eq:augmented_bg_sqrt} & \mathbf{S}^b = \dfrac{1}{\sqrt{n-1}} [\mathbf{x}_1^b - \bar{\mathbf{x}}^b,\dotsb, \mathbf{x}_n^b - \bar{\mathbf{x}}^b] \, ,
\end{align}
\end{subequations}
\end{linenomath*}
where $\bar{\mathbf{x}}^b$ and $\mathbf{S}^b$ are the sample mean and a square root matrix of the sample covariance of $\mathbf{X}^b$, respectively. Here $\bar{\mathbf{x}}^b$ consists of two components, $\bar{\mathbf{\eta}}^b$ and $\bar{\mathbf{\xi}}^b$, which are the sample means of the ensembles of $\mathbf{\Phi}^b$ and $\mathbf{\Xi}^b$, respectively. On the other hand, define
\begin{linenomath*}
\begin{equation} \label{eq: bg_sub_ensemles}
\begin{split}
& \mathbf{S}_{\eta}^b = \dfrac{1}{\sqrt{n-1}} \, [\mathbf{\eta}_1^b - \bar{\mathbf{\eta}}^b, \dotsb, \mathbf{\eta}_n^b - \bar{\mathbf{\eta}}^b ] \, , \\
& \mathbf{S}_{\xi}^b = \dfrac{1}{\sqrt{n-1}} \, [\mathbf{\xi}_1^b - \bar{\mathbf{\xi}}^b, \dotsb, \mathbf{\xi}_n^b - \bar{\mathbf{\xi}}^b ] \, ,
\end{split}
\end{equation}
\end{linenomath*}
then $ \mathbf{S}^b= [(\mathbf{S}_{\eta}^b)^T \,(\mathbf{S}_{\xi}^b)^T ]^T$. Furthermore, let $\mathbf{Y}^b \equiv \{ \mathbf{y}_i^b:\mathbf{y}_i^b =  \mathcal{H} \mathbf{x}_i^b, i = 1, \dotsb, n \}$ be the ensemble of forecasts of the projection of $\mathbf{X}^b$ onto the observation space (projection ensemble for short), then one can construct an $m_y \times n$ matrix (projection matrix for short)
\begin{linenomath*}
\begin{equation} \label{eq:augmented_projected_sqrt}
\mathbf{S}^h = \dfrac{1}{\sqrt{n-1}} \, [\mathbf{y}_1^b - \bar{\mathbf{y}}^b, \dotsb, \mathbf{y}_n^b - \bar{\mathbf{y}}^b ] \, , 
\end{equation}
\end{linenomath*}
where
\begin{linenomath*}
\begin{equation} \label{eq:augmented_prj_mean}
\bar{\mathbf{y}}^b = \dfrac{1}{n} \sum\limits_{i=1}^{n} \mathbf{y}_i^b \, .
\end{equation}
\end{linenomath*}
Similarly, one can also decompose the projection ensemble $\mathbf{Y}^b = \{ \mathbf{y}_i^b: \mathbf{y}_i^b =  \mathcal{H} \mathbf{x}_i^b, i = 1, \dotsb, n \}$ into two parts, $\mathbf{\Phi}^{obv} \equiv \{\mathbf{y}_{\eta,i}^b: \mathbf{y}_{\eta,i}^b = \mathcal{H}_{\eta} \, \mathbf{\eta}_i^b, i = 1, \dotsb, n \}$ and $\mathbf{\Xi}^{obv} \equiv \{\mathbf{y}_{\xi,i}^b: \mathbf{y}_{\xi,i}^b = \mathcal{H}_{\xi} \, \mathbf{\xi}_i^b, i = 1, \dotsb, n \}$, which satisfy $\mathbf{y}_i^b = [(\mathbf{y}_{\eta,i}^b)^T \, (\mathbf{y}_{\xi,i}^b)^T ]^T$ for $i = 1, \dotsb, n$. Let the sample means of $\mathbf{\Phi}^{obv}$ and $\mathbf{\Xi}^{obv}$ be $\bar{\mathbf{y}}_{\eta}^b$ and $\bar{\mathbf{y}}_{\xi}^b$, respectively, such that $\bar{\mathbf{y}}^b = [(\bar{\mathbf{y}}_{\eta}^b)^T \,(\bar{\mathbf{y}}_{\xi}^b)^T]^T$, then the projection matrix in Eq.~(\ref{eq:augmented_projected_sqrt}) can also be decomposed as $\mathbf{S}^h = [(\mathbf{S}_{\eta}^h)^T \, (\mathbf{S}_{\xi}^h)^T ]^T$, where
\begin{linenomath*}
\begin{equation} \label{eq:divided_projected_sqrts}
\begin{split}
& \mathbf{S}_{\eta}^h = \dfrac{1}{\sqrt{n-1}} \, [\mathbf{y}_{\eta,1}^b - \bar{\mathbf{y}}_{\eta}^b, \dotsb, \mathbf{y}_{\eta,n}^b - \bar{\mathbf{y}}_{\eta}^b] \, , \\
& \mathbf{S}_{\xi}^h = \dfrac{1}{\sqrt{n-1}} \, [\mathbf{y}_{\xi,1}^b - \bar{\mathbf{y}}_{\xi}^b, \dotsb, \mathbf{y}_{\xi,n}^b - \bar{\mathbf{y}}_{\xi}^b] \, .
\end{split}
\end{equation}
\end{linenomath*}

\subsection{Implementation of the ETKF in the joint estimation framework} \label{subsec:en_joint}

In the joint estimation framework, the filtering step of the ETKF is given by 
\begin{linenomath*}
\begin{subequations} \label{eq:ETKF_filtering}
\begin{align}
\label{eq:ETKF_mean_update} & \bar{\mathbf{x}}^a = \bar{\mathbf{x}}^b + \mathbf{K} (\mathbf{y} - \mathcal{H} \bar{\mathbf{x}}^b) \, , \\
\label{eq:ETKF_sqrt_update} & \mathbf{S}^a = \mathbf{S}^b \mathbf{T}_{n-1} \mathbf{U} \, , \\
\label{eq:ETKF_KF_gain} & \mathbf{K} =  \mathbf{S}^b (\mathbf{S}^h)^T [\mathbf{S}^h (\mathbf{S}^h)^T + \mathbf{R}]^{-1} \, .
\end{align}
\end{subequations}
\end{linenomath*}
In Eq. (\ref{eq:ETKF_filtering}), $\mathbf{K}$ is the Kalman gain; $\mathbf{T}_{n-1}$ is the $n \times (n-1)$ transform matrix. Roughly speaking, $\mathbf{T}_{n-1}$ is an approximate square root of the matrix $\boldsymbol{\Lambda} = [\mathbf{I}_n + (\mathbf{S}^h)^T \mathbf{R}^{-1} \mathbf{S}^h]^{-1}$ (with $\mathbf{I}_{n}$ being the $n$-dimensional identity matrix), and is constructed based on the $(n-1)$ leading eigenvalues of $\boldsymbol{\Lambda}$ and the associated eigenvectors (see \citealp{Wang-which}); and the $(n-1) \times n$ matrix $\mathbf{U}$ (called centering matrix) satisfies $\mathbf{U} (\mathbf{U})^T = \mathbf{I}_{n-1}$ and $\mathbf{U} \mathbf{1}_n = \mathbf{0}$ \citep{Livings-unbiased,Wang-which}, where $\mathbf{1}_n$ is an $n$-dimensional vector with all its elements being 1. Readers are referred to \citet{Hoteit2002,Wang-which} for the construction of such a centering matrix. Also note that it can be more convenient to use the square root update formula ${\mathbf{S}}^a = \tilde{\mathbf{T}} {\mathbf{S}}^b$, with the transform matrix $\tilde{\mathbf{T}}$ in front of ${\mathbf{S}}^b$, when the ensemble size is larger than the dimension of the observation space \citep{Posselt2012nonlinear}.

With $\bar{\mathbf{x}}^a$ and $\mathbf{S}^a$, the analysis ensemble $ \mathbf{X}^a \equiv \{\mathbf{x}_i^a, i =1 , \dotsb, n \}$ is generated by
\begin{linenomath*}
\begin{equation} \label{eq: augmented_analysis_ensemble}
\mathbf{x}_i^a = \bar{\mathbf{x}}^a + \sqrt{n-1} (\mathbf{S}^a)_i, ~\text{for}~i = 1, \dotsb, n \, ,
\end{equation}
\end{linenomath*}
where $(\mathbf{S}^a)_i$ denotes the $i$-th column of $\mathbf{S}^a$. Propagating $\mathbf{X}^a$ forward, one obtains a background ensemble at the next time step and a new assimilation cycle can begin.

\subsection{Implementation of the ETKF in the divided estimation framework} \label{subsec:en_divided}
In the divided estimation framework, we express all the quantities in the ETKF, e.g., the mean, the square root matrices and the Kalman gain, in terms of some quantities with respect to the sub-systems, such that the divided estimation framework is mathematically equivalent to its joint estimation counterpart. In doing so, the mean update formulae of the ETKF in the divided estimation framework are given by
\begin{linenomath*}
\begin{subequations} \label{eq:divided_mean_update_formulae}
\begin{align}
\label{eq:eta_mean_update} & \bar{\mathbf{\eta}}^a = \bar{\mathbf{\eta}}^b + \mathbf{K}_{11} (\mathbf{y}_{\eta} - \mathcal{H}_{\eta} \, \bar{\mathbf{\eta}}^b) + \mathbf{K}_{12} (\mathbf{y}_{\xi} - \mathcal{H}_{\xi} \, \bar{\mathbf{\xi}}^b) \, , \\
\label{eq:xi_mean_update} & \bar{\mathbf{\xi}}^a = \bar{\mathbf{\xi}}^b + \mathbf{K}_{21} (\mathbf{y}_{\eta} - \mathcal{H}_{\eta} \, \bar{\mathbf{\eta}}^b) + \mathbf{K}_{22} (\mathbf{y}_{\xi} - \mathcal{H}_{\xi} \, \bar{\mathbf{\xi}}^b) \, ,
\end{align}
\end{subequations}
\end{linenomath*}
where
\begin{linenomath*}
\begin{subequations} \label{eq:Kalman_gain_expansion}
\begin{align}
& \mathbf{K}_{11}  = \mathbf{S}_{\eta}^b \mathbf{T}_{\xi} (\mathbf{S}_{\eta}^h \mathbf{T}_{\xi})^T [(\mathbf{S}_{\eta}^h \mathbf{T}_{\xi}) (\mathbf{S}_{\eta}^h \mathbf{T}_{\xi})^T + \mathbf{R}_{\eta}]^{-1} \, , \\
& \mathbf{K}_{12}  = \mathbf{S}_{\eta}^b \mathbf{T}_{\eta} (\mathbf{S}_{\xi}^h \mathbf{T}_{\eta})^T [(\mathbf{S}_{\xi}^h \mathbf{T}_{\eta}) (\mathbf{S}_{\xi}^h \mathbf{T}_{\eta})^T + \mathbf{R}_{\xi}]^{-1} \, , \\
& \mathbf{K}_{21}  = \mathbf{S}_{\xi}^b \mathbf{T}_{\xi} (\mathbf{S}_{\eta}^h \mathbf{T}_{\xi})^T [(\mathbf{S}_{\eta}^h \mathbf{T}_{\xi}) (\mathbf{S}_{\eta}^h \mathbf{T}_{\xi})^T + \mathbf{R}_{\eta}]^{-1} \, , \\
& \mathbf{K}_{22}  = \mathbf{S}_{\xi}^b \mathbf{T}_{\eta} (\mathbf{S}_{\xi}^h \mathbf{T}_{\eta})^T [(\mathbf{S}_{\xi}^h \mathbf{T}_{\eta}) (\mathbf{S}_{\xi}^h \mathbf{T}_{\eta})^T + \mathbf{R}_{\xi}]^{-1} \, ,
\end{align}
\end{subequations}
\end{linenomath*}
with $\mathbf{T}_{\eta}$ and $\mathbf{T}_{\xi}$ being some square root matrices of $[\mathbf{I} + (\mathbf{S}_{\eta}^h)^T \mathbf{R}_{\eta}^{-1} \mathbf{S}_{\eta}^h]^{-1}$ and $[\mathbf{I} + (\mathbf{S}_{\xi}^h)^T \mathbf{R}_{\xi}^{-1} \mathbf{S}_{\xi}^h]^{-1}$, respectively. The derivation of the above formulae is given in Appendix~\ref{sec:appendix A}. 

Based on Eq.~(\ref{eq:ETKF_sqrt_update}), the derivation of the square root update formulae in the divided estimation framework is relatively straightforward. Using the assumption $\mathbf{R} =diag (\mathbf{R}_{\eta}, \mathbf{R}_{\xi})$, one has $(\mathbf{S}^h)^T \mathbf{R}^{-1} \mathbf{S}^h = (\mathbf{S}^h_{\eta})^T \mathbf{R}^{-1}_{\eta} \mathbf{S}^h_{\eta} + (\mathbf{S}^h_{\xi})^T \mathbf{R}^{-1}_{\xi} \mathbf{S}^h_{\xi}$, expressed in terms of the sub-system quantities. Therefore, the transform matrix $\mathbf{T}_{n-1}$ is now constructed based on the $(n-1)$ leading eigenvalues and the corresponding eigenvectors of $[\mathbf{I}_n + (\mathbf{S}^h_{\eta})^T \mathbf{R}^{-1}_{\eta} \mathbf{S}^h_{\eta} + (\mathbf{S}^h_{\xi})^T \mathbf{R}^{-1}_{\xi} \mathbf{S}^h_{\xi}]^{-1}$, and the square root update formulae become
\begin{linenomath*}
\begin{subequations} \label{eq:divided_en_sqrt_update}
\begin{align}
 \label{eq:divided_ensemble_sqrt_update_eta} & \mathbf{S}^a_{\eta} = \mathbf{S}^b_{\eta} \mathbf{T}_{n-1} \mathbf{U} \, , \\
 \label{eq:divided_ensemble_sqrt_update_xi} & \mathbf{S}^a_{\xi} = \mathbf{S}^b_{\xi} \mathbf{T}_{n-1} \mathbf{U} \, ,
\end{align}
\end{subequations}
\end{linenomath*}
with $\mathbf{U}$ being the same $(n-1) \times n$ centering matrix as previously discussed.

Accordingly, the analysis ensembles $\mathbf{\Phi}^a \equiv \{\mathbf{\eta}_i^a, i = 1, \dotsb, n \}$ and $\mathbf{\Xi}^a \equiv \{\mathbf{\xi}_i^a, i = 1, \dotsb, n \}$ are obtained from
\begin{linenomath*}
\begin{subequations} \label{eq: divided_analysis_ensemble}
\begin{align}
 \label{eq: divided_analysis_ensemble_eta} & \mathbf{\eta}_i^a = \bar{\mathbf{\eta}}^a + \sqrt{n-1} (\mathbf{S}^a_{\eta})_i, ~\text{for}~i = 1, \dotsb, n \, , \\
  \label{eq: divided_analysis_ensemble_xi} & \mathbf{\xi}_i^a = \bar{\mathbf{\xi}}^a + \sqrt{n-1} (\mathbf{S}^a_{\xi})_i, ~\text{for}~i = 1, \dotsb, n \, .
\end{align}
\end{subequations}
\end{linenomath*}
Again, by propagating these two ensembles forward through the individual sub-systems, one obtains the background ensembles for the next assimilation cycle.

The mean update formulae Eqs.~(\ref{eq:eta_mean_update},\ref{eq:xi_mean_update}) in the divided estimation framework are similar to that in Eq.~(\ref{eq:ETKF_mean_update}). However, they also exhibit clear differences. For instance, the correction terms in the divided estimation framework, say $\mathbf{K}_{11} (\mathbf{y}_{\eta} - \mathcal{H}_{\eta} \, \bar{\mathbf{\eta}}^b)$ and $\mathbf{K}_{12} (\mathbf{y}_{\xi} - \mathcal{H}_{\xi} \, \bar{\mathbf{\xi}}^b)$ in Eq. (\ref{eq:eta_mean_update}), are associated with some gain matrices, say $\mathbf{K}_{11}$ and $\mathbf{K}_{12}$, that bear different forms from the Kalman gain $\mathbf{K}$ in Eq.~(\ref{eq:ETKF_KF_gain}). There are certain similarities among these gain matrices as well. For instance, if one replaces $\mathbf{S}_{\eta}^b \mathbf{T}_{\xi}$ by $\mathbf{S}_{\eta}^b$ and $\mathbf{S}_{\eta}^h \mathbf{T}_{\xi}$ by $\mathbf{S}_{\eta}^h$, then the gain matrix $\mathbf{K}_{11}$ reduces to the Kalman gain with respect to the sub-system $\eta$. In this sense, the presence of the term $\mathbf{T}_{\xi}$ in $\mathbf{K}_{11}$ reflects the coupling between the sub-systems $\mathbf{\eta}$ and $\mathbf{\xi}$. Similar results can also be found for the other gain matrices $\mathbf{K}_{12}$, $\mathbf{K}_{21}$ and $\mathbf{K}_{22}$. The square root update formula, say Eq.~(\ref{eq:divided_ensemble_sqrt_update_eta}) for the sub-system $\mathbf{\eta}$, has its transform matrix $\mathbf{T}_{n-1}$ as an approximate square root matrix of $[\mathbf{I} + (\mathbf{S}^h_{\eta})^T \mathbf{R}^{-1}_{\eta} \mathbf{S}^h_{\eta} + (\mathbf{S}^h_{\xi})^T \mathbf{R}^{-1}_{\xi} \mathbf{S}^h_{\xi}]^{-1}$, rather than $[\mathbf{I} + (\mathbf{S}^h_{\eta})^T \mathbf{R}^{-1}_{\eta} \mathbf{S}^h_{\eta}]^{-1}$. The extra term $(\mathbf{S}^h_{\xi})^T \mathbf{R}^{-1}_{\xi} \mathbf{S}^h_{\xi}$ also represents the effect of coupling between the sub-systems.

\section{Numerical experiments} \label{sec:experiments}

\subsection{Experiment settings}

We consider the data assimilation problem in a multi-scale Lorenz 96 (ms-L96 hereafter) model \citep[Eqs.~(3.2) and (3.3)]{Lorenz-predictability}, whose governing equations are given by
\begin{linenomath*}
\begin{equation} \label{eq:coupled_L96}
\begin{split}
& \frac{dx_i}{dt} = x_{i-1} (x_{i+1} - x_{i-2}) - x_i + F - \dfrac{hc}{b} \sum\limits_{j=1}^{K} z_{j,i} \, ,  \\
& \frac{dz_{j,i}}{dt} = c b z_{j+1,i} (z_{j-1,i} - z_{j+2,i}) - c z_{j,i} +  \dfrac{hc}{b} x_i\, ,  \\
\end{split}
\end{equation}
\end{linenomath*}
where $i = 1,\dotsb,m$ and $j=1,\dotsb,K$, and $F, c, b, h$ are constant parameters. The state variables $x_i$'s and $z_{j,i}$'s are cyclic as in the Lorenz 96 model \citep{Lorenz-optimal}. For instance, one has $x_{m+1} = x_1; x_0 = x_m; z_{K+1,i} = z_{1,i}; z_{0,i} = z_{K,i}$ etc. In the experiments we let $m=40$, $K=1$, $F=8$, $c=b=10$ and $h=0.8$. This results in a 80-dimensional dynamical system with 40 $x_i$ variables and 40 $z_{1,i}$ variables. In the divided estimation framework the two sub-systems consist of the ordinary differential equations (ODEs) starting with $dx_i/dt$ and $dz_{1,i}/dt$, respectively, i.e., $x_i$ and $z_{1,i}$ play the roles of $\eta$ and $\xi$ in Section \ref{sec:ensemble_filtering}. For convenience, we call the component $x_i$ fast mode (in terms of the rate of state change), and $z_{1,i}$ slow mode, respectively. Fig. \ref{fig:coupledL96_time_series} plots the time series of some state variables in the ms-L96 model.

The dynamical system Eq. (\ref{eq:coupled_L96}) is numerically integrated using the 4th-order Runge-Kutta-Fehlberg (RKF) method \citep{Fehlberg1970classical}, and the system states are collected every $0.05$ time unit (for brevity we call it an integration step). In the experiments we run the system forward in time for 1500 integration steps, and discard the first 500 steps to avoid a spin-up period. In both the joint and divided estimation frameworks, data assimilation starts from step 501 until step 1500. The trajectory during this period is considered as the truth. Synthetic observations are generated by adding Gaussian white noise (with zero mean and unit variance) to the fast mode state variables ${x_1,x_5,x_9,\dotsb,x_{37}}$ and to the slow mode ones ${z_{1,1},z_{1,5},z_{1,9},\dotsb,z_{1,37}}$ (i.e., every 4 state variables and) every 4 integration steps. Therefore, observations are available at 250 out of 1000 integration steps, from 20 out of 80 state variables. For convenience, we re-label the integration step 501 as the first assimilation step. An initial background ensemble with 20 ensemble members is generated by drawing samples from the 80-dimensional multivariate normal distribution $N(\mathbf{0},\mathbf{I}_{80})$ and then adding these samples to the true state at the first assimilation cycle.

In the experiments below we consider an extra possibility, in which the integration of the sub-systems $x_i$ and $z_{1,i}$ in Eq. (\ref{eq:coupled_L96}) is also carried out in a ``divided'' way. This is achieved by temporally treating variables (say $z_{1,i}$'s) as constant parameters in the sub-system (say $x_i$) during the integration, and vice versa. Such a parametrization may incur extra numerical errors during the integration steps. Our main motivation to consider this option is, however, for its potential usefulness in data assimilation practices. For instance, it could be a fast -- although crude, and likely not the best possible -- way to combine earth's sub-system (e.g., ocean, atmosphere etc.) models independently developed by different research groups, and hence increase the re-usability of existing resources. However, it is worthwhile to stress that running the sub-systems separately is not mandatory for the implementation of the divided estimation framework.

Therefore in each experiment below we consider four possible scenarios, which differentiate from each other depending on whether they divide the dynamical system and/or the assimilation scheme. For convenience, we denote these scenarios by (DS-joint,DA-joint), (DS-joint,DA-divided), (DS-divided,DA-joint) and (DS-divided,DA-divided), respectively, where the abbreviations "DS" and "DA" stand for "dynamical system" and "data assimilation", respectively. Here, for instance, "DS-joint" means that the dynamical system is integrated as a whole, and "DA-divided" means that the divided estimation framework is adopted for data assimilation. Other terminologies are interpreted in a similar way.

For illustration, Fig.~\ref{fig:flowchart_ensdkf} outlines the main procedures in the scenario (DS-divided,DA-divided). Starting with an initial ensemble of the coupled system, we split the initial ensemble into two sub-ensembles according to fast and slow modes, and mark them by letters ``F'' and ``S'', respectively. The sub-ensemble ``F'' (``S'') acts as the input state vectors of the fast (slow) mode (denoted by solid arrow lines), and as the input ``parameters'' of the slow (fast) mode (denoted by dotted arrow lines). With incoming observations, the background ensembles of the fast and slow modes are updated to their analysis counterparts as described in Section~\ref{sec:ensemble_filtering}\ref{subsec:en_divided}. Propagating the analysis ensembles forward, one starts a new assimilation cycle, and so on.

Below we compare the performance of the four scenarios through two sets of experiments. In the first set, we conduct the experiment in a plain setting, i.e., without introducing covariance inflation \citep{Anderson-Monte} or localization \citep{Hamill-distance} to the filter. In the second one, covariance inflation and localization are adopted, and the details will be presented later. In all experiments the different scenarios share the same truth, initial ensemble, and observations. For comparison, we use the root mean squared error (RMSE) as a performance measure. For an $m_x$-dimensional system, the RMSE $e_k$ of an analysis $\bar{\mathbf{x}}_k = (\bar{x}_{k,1}, \dotsb, \bar{x}_{k,m_x})^T$ with respect to the truth $\mathbf{x}_k = (x_{k,1}, \dotsb, x_{k,m_x})^T$ at time instant $k$ is defined as
\begin{linenomath*}
\begin{equation} \label{eq:def_of_rmse}
e_k = \dfrac{\Vert \bar{\mathbf{x}}_k - \mathbf{x}_k  \Vert_2}{\sqrt{m_x}} = \sqrt{\dfrac{1}{m_x} \sum\limits_{j=1}^{m_x} (\bar{x}_{k,j} - x_{k,j})^2} ~ .
\end{equation}
\end{linenomath*}

\subsection{Experiment results} \label{subsec:diff_obv}

\subsubsection{Results with the plain setting}

First we investigate whether the joint and divided estimation frameworks yield the same results. To this end, we compare the analyses obtained in both methods by conducting a single update step using identical background ensemble and observations. The experiment is repeated 100 times, each time the background ensemble and observations are drawn at random so that in general they will change over different repetitions. Fig. \ref{fig:coupledL96_first_cycle_diff} shows that the mean and standard deviation (STD) of the differences (in absolute values) between the state variables of the analyses of both estimation frameworks are in the order of $10^{-16}$. Our computations are carried out with MATLAB (version R2012a), in which the numerical precision $\text{eps} = 2.2204 \times 10^{-16}$. This indicates that the tiny differences reported in Fig. \ref{fig:coupledL96_first_cycle_diff} mainly stem from the numerical precision in computations.

Fig. \ref{fig:rmse_plain_setting} depicts the time series of the RMSEs of the estimates obtained in the four different scenarios, (DS-joint,DA-joint), (DS-joint,DA-divided), (DS-divided,DA-joint) and (DS-divided,DA-divided), with a longer time horizon. These four assimilation scenarios have identical initial background ensembles and observations. However, the background ensembles in these four scenarios may (gradually) deviate from each other at subsequent time instants, due to the chaotic nature of the ms-L96 model and the extra parametrization errors in the DS-divided scenarios. 
Therefore, in Fig. \ref{fig:rmse_plain_setting} one can see that, in the DS-joint scenarios, the differences between the estimates from the joint (Panel (a)) and divided (Panel (b)) estimation frameworks are nearly zero during the early assimilation period, but become more substantial over time. Meanwhile, in the DS-divided scenarios, the estimates from either the joint (Panel (c)) or the divided (Panel (d)) estimation framework deviate from those in the (DS-joint,DA-joint) scenario (Panel (a)) more quickly with the extra parametrization errors. 

In terms of estimation accuracy, the time mean RMSE in Panel (a) of Fig. \ref{fig:rmse_plain_setting} is 2.7866. In contrast, the time mean RMSEs in Panels (b-d) are -0.1203 (lower), -0.1649 (lower) and +0.3808 (higher), respectively, relative to that in Panel (a). This seems to suggest that the extra numerical errors due to parametrization are not always harmful. For instance, the time mean RMSE in Panel (c) appears to be the lowest in these four tested scenarios. A possible explanation of this result is discussed later, from the point of view of covariance inflation.

Because the interactions of the forecast and update steps in assimilating the ms-L96 model, it is challenging to obtain an analytic description of the dynamics of the differences between the reference trajectory of the (DS-joint,DA-joint) scenario and those of the (DS-joint,DA-divided), (DS-divided,DA-joint) and (DS-divided,DA-divided) scenarios. For this reason, in what follows we adopt two statistical measures, namely, the boxplot (see the left column of Fig. \ref{fig:diff_box_plot}) and the histogram (see the right column of Fig. \ref{fig:diff_box_plot}), to characterize these differences.  

A boxplot depicts a group of data through their quartiles. In this work the boxplot is adopted to plot the differences at certain time instants. The differences are 80-dimensional vectors, obtained by subtracting the trajectory of the reference scenario (DS-joint,DA-joint) from those of the scenarios (DS-joint,DA-divided), (DS-divided,DA-joint) and (DS-divided,DA-divided) at some particular time instants. A boxplot is used here to indicate the spatial distribution of the 80 elements in a difference vector at a particular time instant. For ease of visualization, we only plot the boxes at time steps $\{1:10:91\}$ and $\{100:100:1000\}$, where $v_{ini} : \Delta v: v_{final}$ stands for an array of scalars that grow from the initial value $v_{ini}$ to the final one $v_{final}$, with an even increment $\Delta v$ each time. Our boxplot setting follows the custom in MATLAB$^{\copyright}$ (version R2012a): On each box, the band inside the box denotes the median, the bottom and top of the box represent the 25th and 75th percentiles, the ends of the whiskers indicate the extension of the data that are considered non-outliers, while outliers are marked individually as asterisks in Fig. \ref{fig:diff_box_plot}. Note that, in the (DS-joint,DA-divided) scenario, because the differences from the reference trajectory are very tiny at the early assimilation stage, the boxes appear to collapse during this period (e.g., from time steps 1 to 91), which is consistent with the results in Fig. \ref{fig:etkf_noSysSplit_estSplit}. As time moves forward, the trajectory of the (DS-joint,DA-divided) scenario gradually deviate from the reference. Therefore, as indicated in Fig. \ref{fig:boxplot_sysSplit_estSplit_trajectory_Diff_DS-joint_DA-divided}, the spreads of the differences become larger from time step 200 on, compared to those at earlier time steps. In addition, more outliers (asterisks) are seen after time step 200, while the medians of the differences appear to remain close to zero at all time steps. Similar phenomena can also be observed in the (DS-divided,DA-joint) and (DS-divided,DA-divided) scenarios, except that the periods in which the boxplots collapse are much shorter compared to that in the (DS-joint,DA-divided) scenario, which is also consistent with the results in Fig. \ref{fig:rmse_plain_setting}.  

The histogram is also used here to depict the distribution of an element in a difference vector during the whole assimilation time window. In the right column of Fig. \ref{fig:diff_box_plot} we show the $20$th and $60$th elements, which correspond to the trajectory differences in the state variables $x_{20}$ and $z_{1,20}$, respectively, in the scenarios (DS-joint,DA-divided), (DS-divided,DA-joint) and (DS-divided,DA-divided). In the (DS-joint,DA-divided) scenario, the histogram of the differences in state variable $x_{20}$ appears to have a single peak at zero, while its support is inside the interval $[-15 ~ 15]$. The histogram of the differences in state variable $z_{1,20}$ also has a single peak at zero, but its support is narrower, being inside the interval $[-1.5 ~ 1.5]$ instead. Similar phenomena are also observed in the (DS-divided,DA-joint) and (DS-divided,DA-divided) scenarios, although the heights of the peaks tend to be lower, and the corresponding supports tend to be wider. 

Overall, the results in Figs. \ref{fig:rmse_plain_setting} and \ref{fig:diff_box_plot} seem to suggest that the trajectories of the (DS-joint,DA-divided), (DS-divided,DA-joint) and (DS-divided,DA-divided) scenarios tend to oscillate around the reference trajectory of the (DS-joint,DA-joint) scenario, although they may also substantially deviate from the reference one at many time instants.             

\subsubsection{Results with both covariance inflation and localization}
Covariance inflation \citep{Anderson-Monte} and localization \citep{Hamill-distance} are two important auxiliary techniques that can be used to improve the performance of an EnKF. Since the EnKF is a Monte Carlo implementation of the Kalman filter, when the ensemble size is relatively small, certain issues may arise, including, for instance, systematic underestimation of the variances of state variables, overestimation of the correlations of different state variables and rank deficiency in the sample error covariance matrix.

Covariance inflation \citep{Anderson-Monte} is introduced to tackle the variance underestimation problem by artificially increasing the sample error covariance to some extent. In relation to the results in the previous experiment, one possible explanation of the result there is that the extra numerical errors due to parametrization may have acted as some additive noise in the dynamical model, which is not always bad for a filter's performance. Indeed, as has been reported in some earlier works, e.g., \citet{Gordon1993,Hamill2011-what}, introducing some artificial noise to the dynamical model may improve filter performance. In the context of EnKF, this may be considered as an alternative form of covariance inflation \citep{Hamill2011-what}, which may enhance the robustness of the filter from the point of view of $H_{\infty}$ filtering theory \citep{Luo2011_EnLHF,Altaf2013-improving,Triantafyllou2012-assessing}. One may also introduce artificial noise in a more sophisticated way, e.g., through a certain nonlinear regression model, such that the statistical effect of the regression model mimics that of the dynamical model \citep{Harlim2014ensemble}.   

How to optimally conduct covariance inflation is an ongoing research topic in the data assimilation community. Some recent developments include, for example, adaptive covariance inflation techniques (see, for example, \citealp{Anderson2007,Anderson2009}) and covariance inflation from the point of view of residual nudging \citep{Luo2014-efficient,luo2013-covariance,Luo2012-residual}, among many others. For our purpose here, it appears sufficient to conduct covariance inflation by simply multiplying the analysis sample error covariance by a factor $\delta^2$ ($\delta \geq 1$), as originally proposed in \citet{Anderson-Monte}. The values of $\delta$ in the experiment are $\{1:0.05:1.3\}$.

Covariance localization \citep{Hamill-distance} is adopted to deal with the overestimation of the correlations and rank deficiency. In practice, different methods are proposed to conduct localization, for examples, see \citet{Anderson2007,Anderson2009,clayton2012operational,kuhl2013comparison,wang2007comparison}. In our experiments localization is directly applied to the gain matrices. We assume that $z_{1,i}$ and $x_i$ are located at the same grid point $i$. Covariance localization thus follows the settings in \citet{Anderson2007}, in which a parameter $l_c$, called half-width (or length scale of localization), controls the degree of correlation tapering. We use the same half-width for the fast and slow components of the ms-L96 model, with $l_c$ being chosen from the set $\{0.1 :0.2 :0.9 \}$. In general, for both the joint and divided estimation frameworks, one may use different half-widths for different components (e.g., ocean and atmosphere) of a coupled system. In such circumstances, it could be more efficient to use an adaptive localization approach (for examples, see \citealp{bishop2007flow,bishop2009ensembleP1,bishop2009ensembleP2,bishop2011adaptive}).

We investigate the filter performance in the aforementioned four scenarios by combining different values of the inflation factor $\delta$ and the half-width $l_c$. The corresponding results, in terms of time mean RMSEs (the averages of the RMSEs over the assimilation time window) are reported in Fig. \ref{fig:rmse_lc_delta}. In the experiments, the filters' performance is improved in most of the cases, in comparison with the results in Fig. \ref{fig:rmse_plain_setting}. In Fig. \ref{fig:rmse_lc_delta}, the best filter performance is obtained with $l_c \approx 0.7$, while with localization, covariance inflation does not seem to help improve the estimation accuracy\footnote{When covariance localization is excluded, inflation may improve the filters' performance (results not shown).}, similar to the findings of \citet{penny2013hybrid}. The above results, however, may strongly depend on the experimental settings. For instance, in the context of the hybrid local ETKF, \citet{penny2013hybrid} found that the best filter performance is achieved at relatively small $l_c$ values (e.g., $\approx 0.2$). 


Fig. \ref{fig:rmse_lc_delta} also indicates that, for a given model integration scenario (either DS-joint or DS-divided), the joint and divided estimation frameworks yield very close results. On the other hand, for a given estimation framework (either DA-joint or DA-divided), integrating the sub-systems separately tends to deteriorate filter performance. In general the performance deterioration is not severe, less than $10\%$ in all cases with the same values of $\delta$ and $l_c$. 

\section{Two extensions from the practical point of view} \label{sec:extensions}

In this section we present two extensions of the aforementioned frameworks. These are largely motivated by the current status and challenges of conducting data assimilation in coupled ocean-atmosphere models \citep{DAwhitepaper2013}. These two extensions are illustrated within the (DS-divided,DA-divided) scenario. The extensions to the other scenarios can be implemented in a similar way. 

\subsection{Different ensemble sizes in the sub-systems}
Here we consider the possibility of running the filter with different ensemble sizes in the fast and slow modes. This may be considered as an example in which one wants to gain certain computational efficiency by running fewer ensemble members in one of the sub-systems, but possibly at the cost of certain loss of accuracy. 
To this end, let the ensemble sizes of the fast and slow modes be $n_f$ and $n_s$, respectively. In the experiments, we consider four different cases, with $(n_f = 20, n_s = 20)$, $(n_f = 20, n_s = 15)$, $(n_f = 15, n_s = 20)$ and $(n_f = 15, n_s = 15)$, respectively, at the prediction step, and the targeted ensemble size is 20 for both modes at the filtering step. To apply the filter update formulae, the ensemble sizes of both modes should be equal. Therefore dimension mismatch will arise when $n_f \neq n_s$. This issue is addressed through a conditional sampling scheme discussed in the supplementary material.  

In each of the above cases, we investigate the filter's performance when (a) neither covariance inflation nor covariance localization is applied (the plain setting); and (b) both covariance inflation and covariance localization are adopted. In the setting (b), the covariance inflation factor is $1.15$ for both the fast and slow modes, and the half-width for covariance localization is $0.75$.

Fig. \ref{fig:rmse_ensize} plots the time series of the RMSEs for the above four different cases. In each case, when the filter is equipped with both covariance inflation and localization, its time mean RMSE tends to be lower than that of the plain setting (with neither inflation nor localization). On the other hand, if one takes the case $(n_f = 20, n_s = 20)$ with both covariance inflation and localization as the reference, then it is clear that reducing the ensemble size of either the fast or slow mode degrades the filter performance in terms of RMSE. Also, comparing Figs. \ref{fig:L96_ensize_20_15} and \ref{fig:L96_ensize_15_20}, one can see that reducing the ensemble size of the fast mode appears to have a larger (negative) impact than reducing the ensemble size of the slow one, which may be because the fast mode appears to dominate the dynamics of the ms-L96 model (see Fig. \ref{fig:lt_statistics} later, also the similar results in \citealp{Hoteit2004-adaptively}). On the other hand, comparing Figs. \ref{fig:L96_ensize_15_20} and \ref{fig:L96_ensize_15_15}, it seems better to simply reduce the ensemble sizes of both the fast and slow modes, in contrast to the case that reduces the ensemble size of the fast mode only. This may also be because the fast mode is the dominant part to the dynamics of the ms-L96 model, therefore the extra errors due to the sampling scheme may be significant to the filter performance. However, a comparison between Figs. \ref{fig:L96_ensize_20_15} and \ref{fig:L96_ensize_15_15} suggests that if one only reduces the ensemble size of the slow mode, then the filter performance can be better than that resulting from reducing the ensemble sizes of both modes. Similar results are also observed with the plain setting, except that with the plain setting, the case $(n_f = 20, n_s = 15)$ seems to perform slightly better than the one with $(n_f = 20, n_s = 20)$.


\subsection{Incorporating the ensemble optimal interpolation into the divided estimation framework} \label{subsec:EnOI}
If one sub-system of the coupled model (e.g., the ocean in the coupled ocean-atmosphere model) exhibits relatively slow changes, then it may be reasonable to assume that this sub-system has an (almost) constant background covariance over a short assimilation time window \citep{Hoteit2002} \footnote{In the context of meteorological applications, the extension described here mainly targets short-term (e.g., sub-seasonal) time scales, while for seasonal or longer time scale applications (e.g., climate studies), the small-variation assumption (e.g., in the ocean component) may not be valid.}. As a result, optimal interpolation (OI, see, for example, \citealp{Cooper1996}) could be a reasonable assimilation scheme for such a slow-varying sub-system model, due to its simplicity in implementation and significant savings in computational cost. The ensemble optimal interpolation (EnOI, see, for example, \citealp{Counillon2009}) is an ensemble implementation of the OI scheme. It has an update step similar to that of the EnKF, but computes the associated background covariance (or square root matrix) based on a ``historical'' ensemble \citep{Counillon2009}. At the prediction step, the EnOI only propagates the analysis mean forward to obtain a background mean at the next assimilation cycle. This is computationally much cheaper than propagating the whole analysis ensemble forward as in the EnKF, hence appears attractive for certain applications (e.g., oceanography, see \citealp{Hoteit2002,DAwhitepaper2013}).

Here we consider the possibility to tailor the divided estimation framework so as to incorporate the EnOI into one of the sub-systems. Such a modification is largely motivated by the current status and challenges of operational data assimilation in coupled ocean-atmosphere models, in which, due to the limitations in computational resource, one may use OI or 3D-Var (or their ensemble implementations) for the ocean model, and a more sophisticated scheme such as 4D-Var or EnKF for the atmosphere model. Therefore combining these different assimilation systems becomes a challenge in practice \citep{DAwhitepaper2013}.

In our investigation below, to incorporate the EnOI into the divided estimation framework, some modifications are introduced as follows: (a) At the prediction step, the slow mode only propagates forward the analysis mean of the corresponding sub-ensemble, and uses the analysis mean with respect to the fast mode as the ``parameters'' in the numerical integrations of the slow mode. On the other hand, the fast mode propagates forward the corresponding analysis sub-ensemble (updated through Eqs. (\ref{eq:divided_en_sqrt_update}) and (\ref{eq: divided_analysis_ensemble})), and uses the update of the ``historical'' ensemble (also through Eqs. (\ref{eq:divided_en_sqrt_update}) and (\ref{eq: divided_analysis_ensemble})) of the slow mode as the ``parameters'' in the numerical integrations of the fast mode; (b) At the filtering step, the background sub-ensemble of the fast mode is the propagation of the analysis sub-ensemble from the previous assimilation cycle, while the background sub-ensemble of the slow mode is the ``historical'' ensemble generated by drawing a specified number of samples from a Gaussian distribution whose mean and covariance are equal to the ``climatological'' mean and covariance of the slow mode, respectively. This ``historical'' ensemble is produced once for all, and does not change over the assimilation window. However, at each assimilation cycle, when a new observation is available, the ``historical'' ensemble is updated according to Eqs. (\ref{eq:divided_en_sqrt_update}) and (\ref{eq: divided_analysis_ensemble}), and is used as the ``parameters'' of the fast mode. 
In doing so, the cross-covariance between the ``historical'' ensemble of the slow mode and the flow-dependent sub-ensemble of the fast mode may not accurately represent the true correlations between both modes.

To generate the ``historical'' ensemble of the slow mode, we run the ms-L96 model forward in time for 100,000 integration steps, with the step size being 0.05. The ``climatological'' statistics are then taken as the temporal mean and covariance of the generated trajectory. Fig. \ref{fig:lt_statistics} shows the values of the ``climatological'' means and the eigenvalues of the ``climatological'' covariances of the fast and slow modes. These results suggest that the fast mode dominates the slow one in magnitudes, consistent with the results in Fig. \ref{fig:coupledL96_time_series}.

In the experiments below, the ensemble sizes of the fast and slow modes are both 20. For distinction, hereafter we refer to the extended assimilation scheme with the EnOI as "DA-divided-exEnOI", and that without the EnOI as "DA-divided". We also consider two settings: In the plain setting neither covariance inflation nor localization is conducted, while in the other setting both auxiliary techniques are applied, with the inflation factor being 1.15 for the fast and slow modes, and the half-with being 0.7.

Fig. \ref{fig:EnOI} plots the time series of the RMSEs for the DA-divided and DA-divided-exEnOI. When neither covariance inflation nor localization is adopted, the magnitudes of the trajectories of DA-divided and DA-divided-exEnOI are comparable at many time instants, although substantial differences are also spotted in some cases (e.g., the interval between time steps 100 and 200). On the other hand, when covariance inflation and localization are applied, both DA-divided and DA-divided-exEnOI schemes tend to yield lower time mean RMSEs. In addition, with covariance inflation and localization, the difference (in time mean RMSE) between DA-divided and DA-divided-exEnOI is narrowed from around 0.06 to around 0.01. Although the relative performance of the DA-divided and DA-divided-exEnOI schemes may in general change from case to case, the above experiment suggests -- at least for the ms-L96 model -- the potential of incorporating the EnOI into the divided estimation framework to reduce the computational cost. 

\section{Discussion and conclusion} \label{sec:conclusion}

We consider the data assimilation problem in coupled systems composed of two sub-systems. A straightforward method to tackle this problem is to augment the state vectors of the sub-systems. In contrast, the divided estimation framework re-expresses the update formulae in the joint estimation framework in terms of some quantities with respect to the sub-systems themselves. We also consider the option of running the sub-systems separately, which may bring flexibility and efficiency to data assimilation practices in certain situations, but possibly at the cost of larger discretization errors during model integrations.

We use a multi-scale Lorenz 96 model to evaluate the performance of four different data assimilation scenarios, combining different options of joint/divided sub-systems and joint/divided estimation frameworks. In addition, we also consider two possible extensions that may be relevant for certain coupled data assimilation problems. The experiment results suggest that, (a) with identical background ensemble and observation, the joint and divided estimation frameworks yield the same estimate within the machine's numerical precision; (b) running the sub-systems separately may bring in extra flexibility in practice, but at the cost of reduced estimation accuracy in certain circumstances; and (c) for the approximations used in the extension schemes of Section~\ref{sec:extensions}, provided that the assimilation schemes are properly configured, one might still obtain reasonable estimates, especially when both covariance inflation and localization are applied.

The current work mainly services as a proof-of-concept study. In real applications, for instance, data assimilation in coupled ocean-atmosphere general circulation models (OAGCM), model balance and the generation of initial background ensemble are among the issues that require special attention \citep{saha2013ncep,zhang2007system}. Additional challenges (e.g., different time scales between ocean and atmosphere components) may also arise when coupled data assimilation is extended to longer time scales (e.g., in the context of climate studies). In this case, certain configurations in the current work may need to be modified, including (but not limited to), for instance, the way to generate the initial background ensemble and to conduct the conditional sampling (supplementary material). This study may be considered as a complement to some existing works in the literature (e.g., \citealp{zhang2007system}), in terms of the data assimilation schemes in use. In light of the mathematical equivalence between the joint and divided estimation frameworks, we envision that existing techniques (see, for example, \citealp{saha2013ncep,zhang2007system}) and their future developments used to tackle the aforementioned challenges can also be applied in a similar way within the divided estimation framework.   

One may also extend the present work to the situations where the coupled system consists of more than two components. This extension may be of interest in certain situations, for instance, when the interactions of land, ocean and atmosphere are in consideration, or when the domain of a global model is divided into a number of sub-domains such that data assimilation is conducted in a set of regional models, similar to the scenario considered in the local ensemble Kalman filter \citep{Ott-local}. In such cases, the corresponding update formulae may become more complicated when adopting the divided estimation framework. This topic will be investigated in the future.

\section*{Acknowledgement}
We would like to thank three reviewers for their constructive comments and suggestions that significantly improved the presentation and quality of the work. This study was funded by King Abdullah University of Science and Technology (KAUST). The first author would also like to thank the IRIS/CIPR cooperative research project ``Integrated Workflow and Realistic Geology'' which is funded by industry partners ConocoPhillips, Eni,  Petrobras, Statoil, and Total, as well as the Research Council of Norway (PETROMAKS), for partial financial support. 

\appendix
\setcounter{section}{0}
\setcounter{equation}{0}
\renewcommand{\thesection}{\Alph{section}}
\renewcommand{\theequation}{\thesection.\arabic{equation}}

\section{Gain matrices in the divided estimation framework} \label{sec:appendix A}

In the divided estimation framework, the most cumbersome part lies in the expansion of the Kalman gain $\mathbf{K}$ in Eq.~(\ref{eq:ETKF_KF_gain}). Here we split the deduction into a few steps. First of all, we compute the component $\mathbf{S}^b (\mathbf{S}^h)^T$, which reads
\begin{linenomath*}
\begin{equation} \label{eq:divided_K_term1}
\mathbf{S}^b (\mathbf{S}^h)^T =
\begin{pmatrix}
\mathbf{S}_{\eta}^b (\mathbf{S}_{\eta}^h)^T & \mathbf{S}_{\eta}^b (\mathbf{S}_{\xi}^h)^T \\
\mathbf{S}_{\xi}^b (\mathbf{S}_{\eta}^h)^T & \mathbf{S}_{\xi}^b (\mathbf{S}_{\xi}^h)^T \\
\end{pmatrix} \, .
\end{equation}
\end{linenomath*}
Next, we consider the component $[\mathbf{S}^h(\mathbf{S}^h)^T + \mathbf{R}]^{-1}$, which can be expanded as
\begin{linenomath*}
\begin{equation} \label{eq:divided_K_term2}
[\mathbf{S}^h(\mathbf{S}^h)^T + \mathbf{R}]^{-1} =
\begin{pmatrix}
\mathbf{S}_{\eta}^h (\mathbf{S}_{\eta}^h)^T + \mathbf{R}_{\eta} & \mathbf{S}_{\eta}^h (\mathbf{S}_{\xi}^h)^T \\
\mathbf{S}_{\xi}^h (\mathbf{S}_{\eta}^h)^T & \mathbf{S}_{\xi}^h (\mathbf{S}_{\xi}^h)^T + \mathbf{R}_{\xi} \\
\end{pmatrix} ^{-1} \, .
\end{equation}
\end{linenomath*}
Applying the matrix inversion lemma \cite[p.~11]{Simon2006} on the right hand side of Eq.~(\ref{eq:divided_K_term2}), we have
\begin{linenomath*}
\begin{equation} \label{eq:divided_K_inversion_lemma}
[\mathbf{S}^h(\mathbf{S}^h)^T + \mathbf{R}]^{-1} =
\begin{pmatrix}
\mathbf{C}_{\eta}^{-1} & - \mathbf{A}_{\eta \xi} \mathbf{C}_{\xi}^{-1} \\
-\mathbf{A}_{\xi \eta} \mathbf{C}_{\eta}^{-1} & \mathbf{C}_{\xi}^{-1} \\
\end{pmatrix}
 \, ,
\end{equation}
\end{linenomath*}
where
\begin{linenomath*}
\begin{equation}
\begin{split}
& \mathbf{A}_{\eta \xi} = [\mathbf{S}_{\eta}^h (\mathbf{S}_{\eta}^h)^T + \mathbf{R}_{\eta}]^{-1}[\mathbf{S}_{\eta}^h (\mathbf{S}_{\xi}^h)^T] \, , \\
& \mathbf{A}_{\xi \eta} = [\mathbf{S}_{\xi}^h (\mathbf{S}_{\xi}^h)^T + \mathbf{R}_{\xi}]^{-1}[\mathbf{S}_{\xi}^h (\mathbf{S}_{\eta}^h)^T] \, ,
\end{split}
\end{equation}
\end{linenomath*}
and
\begin{linenomath*}
\begin{equation} \label{eq:divided_schur_eta}
\begin{split}
\mathbf{C}_{\eta} & = \mathbf{S}_{\eta}^h (\mathbf{S}_{\eta}^h)^T + \mathbf{R}_{\eta} - [\mathbf{S}_{\eta}^h (\mathbf{S}_{\xi}^h)^T] \mathbf{A}_{\xi \eta} \\
& = \mathbf{R}_{\eta} + \mathbf{S}_{\eta}^h [\mathbf{I} - (\mathbf{S}_{\xi}^h)^T [\mathbf{S}_{\xi}^h (\mathbf{S}_{\xi}^h)^T + \mathbf{R}_{\xi}]^{-1} \mathbf{S}_{\xi}^h] (\mathbf{S}_{\eta}^h)^T \\
& = \mathbf{R}_{\eta} + \mathbf{S}_{\eta}^h [\mathbf{I} + (\mathbf{S}_{\xi}^h)^T \mathbf{R}_{\xi}^{-1} \mathbf{S}_{\xi}^h]^{-1} (\mathbf{S}_{\eta}^h)^T \\
& = \mathbf{R}_{\eta} + (\mathbf{S}_{\eta}^h \mathbf{T}_{\xi}) (\mathbf{S}_{\eta}^h \mathbf{T}_{\xi})^T \, .\\
\end{split}
\end{equation}
\end{linenomath*}
The equality between the second and third lines of Eq.~(\ref{eq:divided_schur_eta}) is derived based on the Sherman--Morrison--Woodbury identity \citep{sherman1950adjustment} such that
\begin{linenomath*}
\begin{equation} \label{eq:mtx_identity_1}
(\mathbf{I} + \mathbf{M}^T \mathbf{R}^{-1} \mathbf{M})^{-1} = \mathbf{I} - \mathbf{M}^T ( \mathbf{M}\mathbf{M}^T+ \mathbf{R})^{-1} \mathbf{M} \, .
\end{equation}
\end{linenomath*}
In the last line of Eq.~(\ref{eq:divided_schur_eta}), $\mathbf{T}_{\xi}$ is a square root of $[\mathbf{I} + (\mathbf{S}_{\xi}^h)^T \mathbf{R}_{\xi}^{-1} \mathbf{S}_{\xi}^h]^{-1}$, and is equivalent to the transform matrix of the ETKF, with respect to the sub-system $\mathbf{\xi}$ \citep{Bishop-adaptive}. Similarly, we have
\begin{linenomath*}
\begin{equation} \label{eq:divided_schur_xi}
\mathbf{C}_{\xi} = \mathbf{R}_{\xi} + (\mathbf{S}_{\xi}^h \mathbf{T}_{\eta}) (\mathbf{S}_{\xi}^h \mathbf{T}_{\eta})^T \, ,
\end{equation}
\end{linenomath*}
with $\mathbf{T}_{\eta}$ being a square root of $[\mathbf{I} + (\mathbf{S}_{\eta}^h)^T \mathbf{R}_{\eta}^{-1} \mathbf{S}_{\eta}^h]^{-1}$.

Combining Eqs.~(\ref{eq:ETKF_KF_gain}), (\ref{eq:divided_K_term1}) and (\ref{eq:divided_K_inversion_lemma}), and with some algebra, we obtain the Kalman gain
\begin{linenomath*}
\begin{equation}
\mathbf{K} = \begin{pmatrix}
\mathbf{K}_{11} & \mathbf{K}_{12}\\
\mathbf{K}_{21} & \mathbf{K}_{22}\\
\end{pmatrix}
 \, ,
\end{equation}
\end{linenomath*}
where
\begin{linenomath*}
\begin{equation} \label{eq:Kalman_gain_11}
\begin{split}
\mathbf{K}_{11} & = [ \mathbf{S}_{\eta}^b (\mathbf{S}_{\eta}^h)^T - \mathbf{S}_{\eta}^b (\mathbf{S}_{\xi}^h)^T [\mathbf{S}_{\xi}^h (\mathbf{S}_{\xi}^h)^T + \mathbf{R}_{\xi}]^{-1}\, \mathbf{S}_{\xi}^h (\mathbf{S}_{\eta}^h)^T] \mathbf{C}_{\eta}^{-1} \\
& = \mathbf{S}_{\eta}^b \mathbf{T}_{\xi} (\mathbf{S}_{\eta}^h \mathbf{T}_{\xi})^T [(\mathbf{S}_{\eta}^h \mathbf{T}_{\xi}) (\mathbf{S}_{\eta}^h \mathbf{T}_{\xi})^T + \mathbf{R}_{\eta}]^{-1} \, .\\
\end{split}
\end{equation}
\end{linenomath*}
The deduction of the last line of Eq.~(\ref{eq:Kalman_gain_11}) is similar to that in Eq.~(\ref{eq:divided_schur_eta}), and hence we omit the details. The other elements of $\mathbf{K}$ can be obtained in a similar way and are summarized in Eq. (\ref{eq:Kalman_gain_expansion}).
\bibliographystyle{ametsoc}
\bibliography{references}

\clearpage
\renewcommand{\thefigure}{\arabic{figure}}

\clearpage
\begin{figure*} 
\centering
\includegraphics[width=0.8\textwidth]{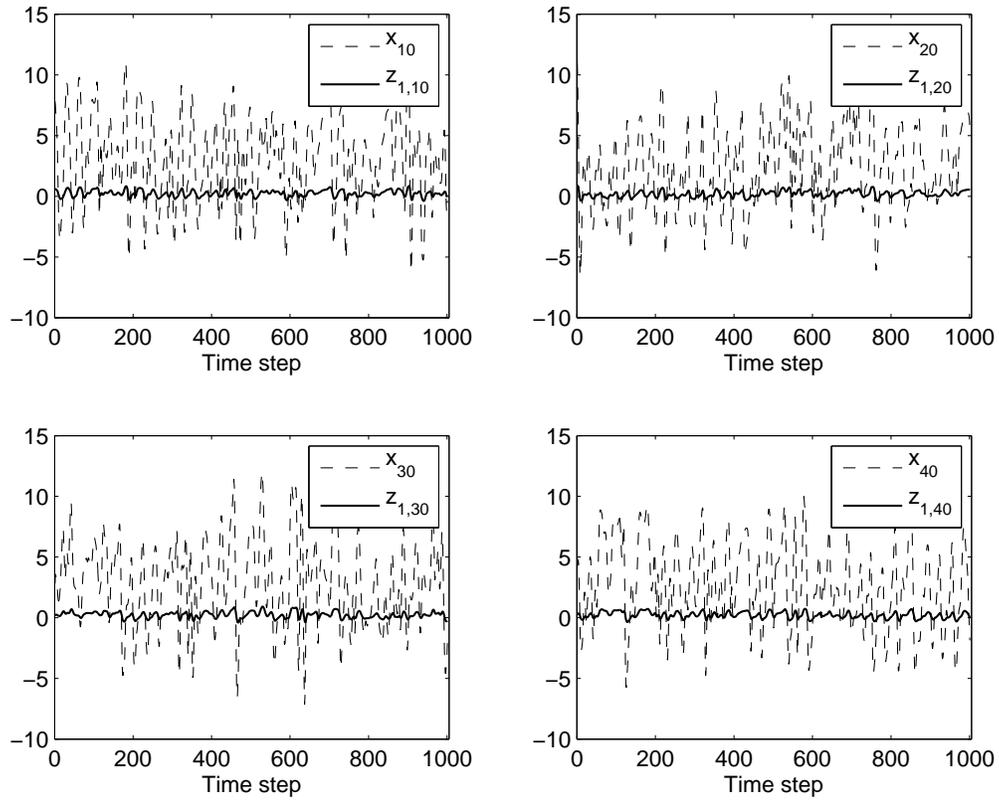}
\caption{ \label{fig:coupledL96_time_series} Time series of some state variables in the multi-scale Lorenz 96 model.}
\end{figure*}

\clearpage
\begin{figure*} 
\centering
\includegraphics[width=0.8\textwidth]{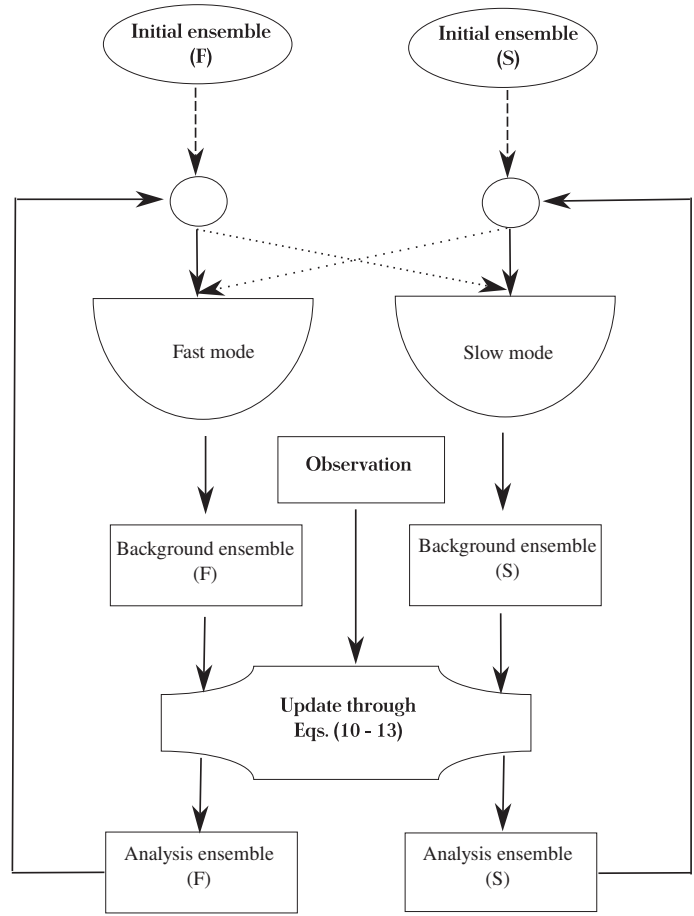}
\caption{ \label{fig:flowchart_ensdkf} Flow chart of the procedures in the (DS-divided, DA-divided) scenario in assimilating the multi-scale Lorenz 96 system.}
\end{figure*}

\clearpage
\begin{figure*} 
\centering

	\subfigure[Mean of absolute differences]{
	\label{fig:mean_abs_diff}	
	\includegraphics[width = 0.45\textwidth] {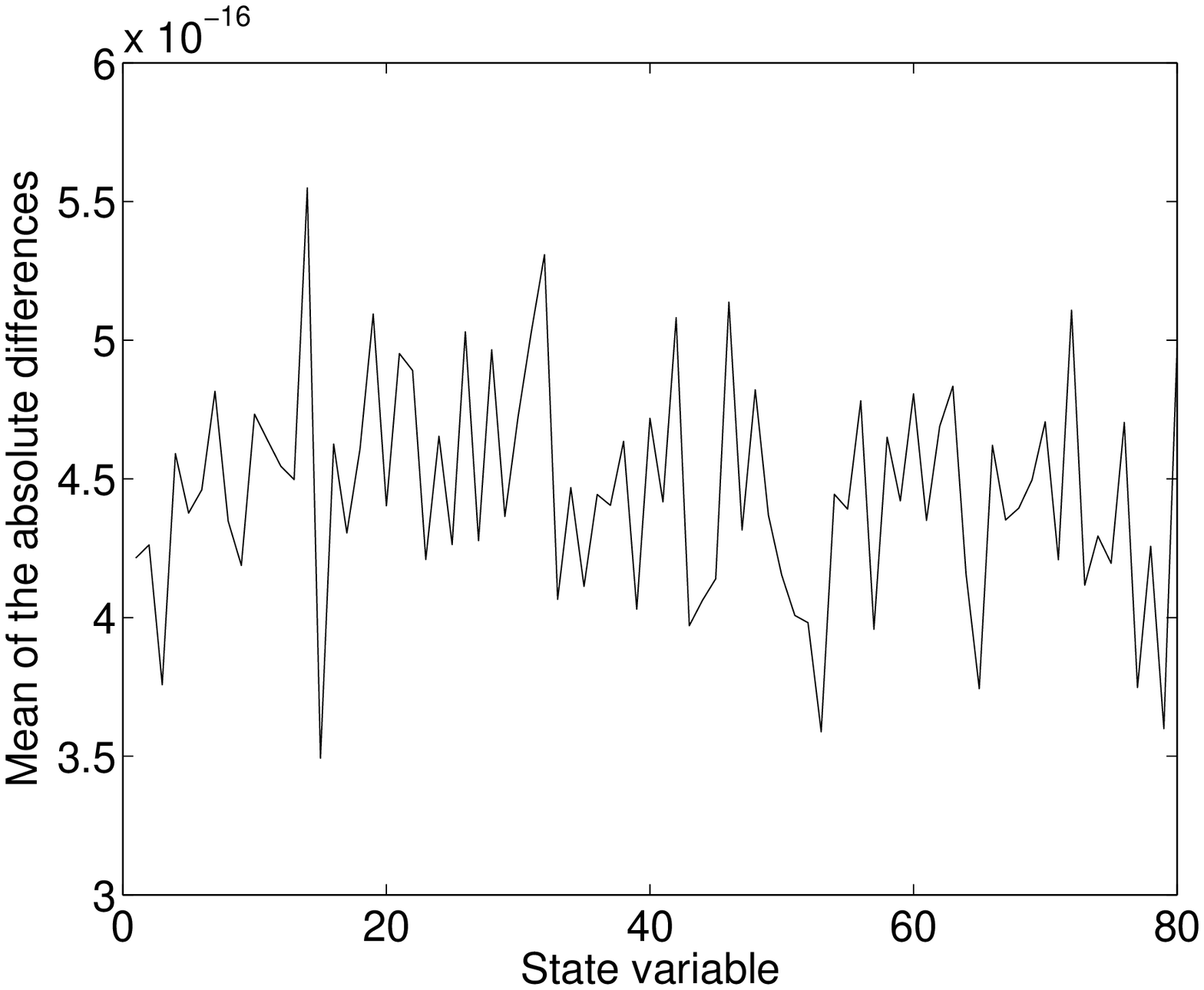}
     }
	\subfigure[STD of absolute differences]{
	\label{fig:std_diff}	
	\includegraphics[width = 0.45\textwidth] {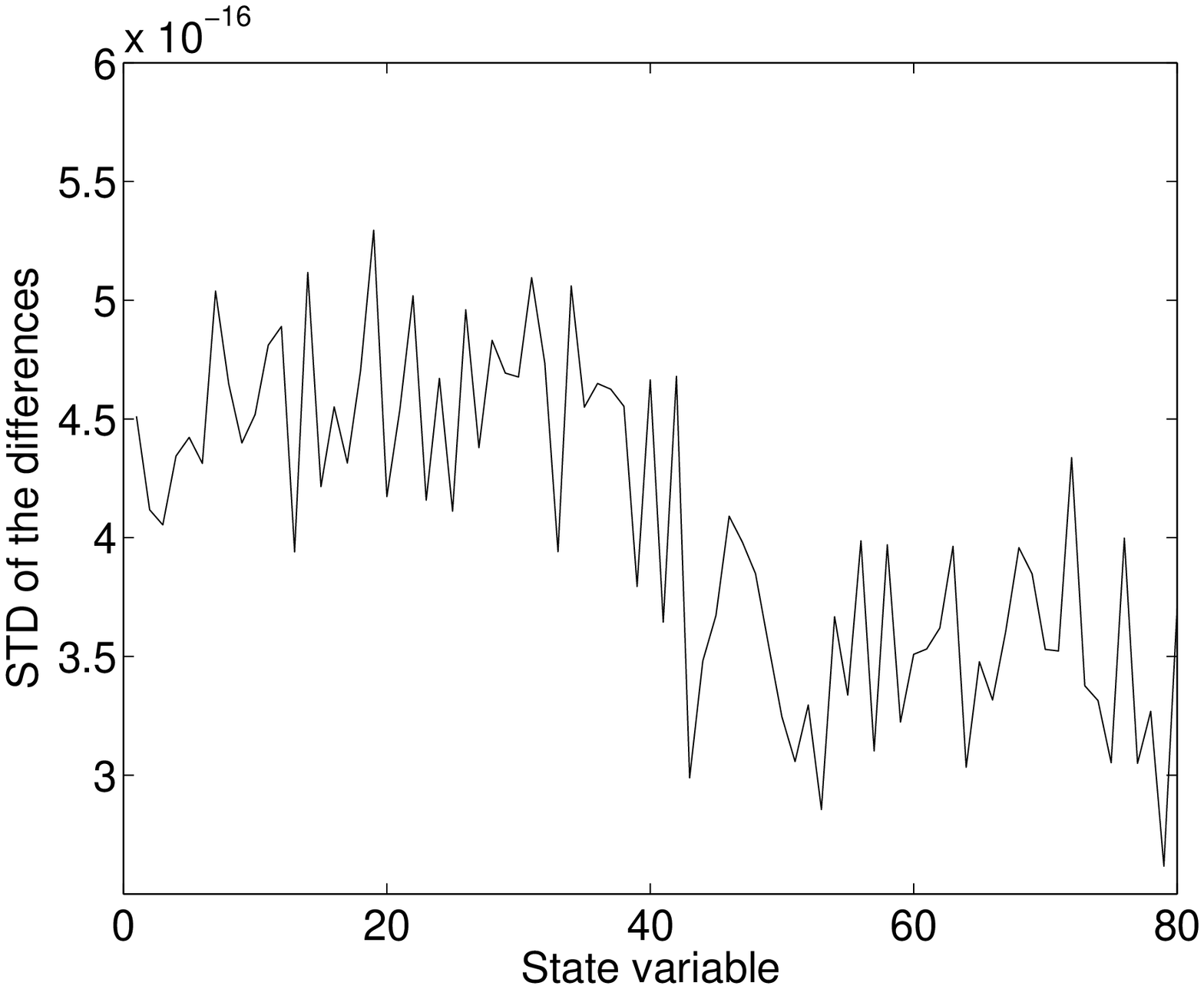}
     } 

\caption{ \label{fig:coupledL96_first_cycle_diff} A comparison of the analyses of joint and divided estimation frameworks with a single update. The experiment is repeated 100 times. In each repetition, the background ensemble and the observation are drawn at random (so that in general they will change over different repetitions), and in each repetition, the joint and divided estimation frameworks share the same background ensemble and observation. Panel (a): Mean value (over 100 repetitions) of the absolute differences between the analyses of the joint and divided estimation frameworks in each state variable. Panel (b): Corresponding standard deviation (STD) of the absolute differences.}
\end{figure*}

\clearpage
\begin{figure*} 
\centering
	\subfigure[Scenario (DS-joint,DA-joint)]{
	\label{fig:etkf_noSysSplit_noEstSplit}	
	\includegraphics[width = 0.45\textwidth] {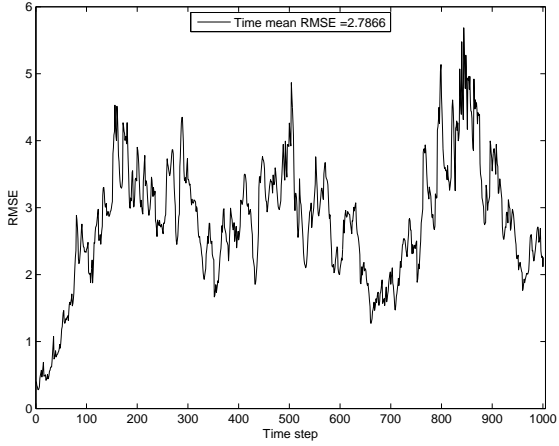}
     }
	\subfigure[Scenario (DS-joint,DA-divided)]{
	\label{fig:etkf_noSysSplit_estSplit}	
	\includegraphics[width = 0.45\textwidth] {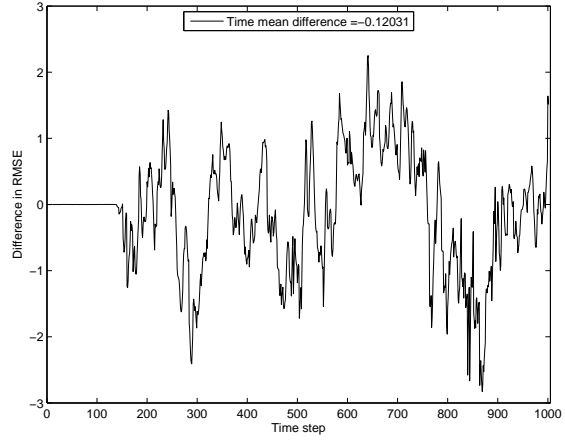}
     } \hspace{-0.0cm}
	\subfigure[Scenario (DS-divided,DA-joint)]{
	\label{fig:etkf_sysSplit_noEstSplit}	
	\includegraphics[width = 0.45\textwidth] {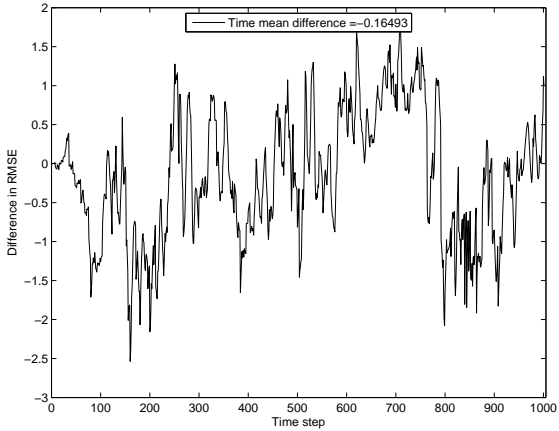}
     }
	\subfigure[Scenario (DS-divided,DA-divided)]{
	\label{fig:etkf_sysSplit_estSplit}	
	\includegraphics[width = 0.45\textwidth] {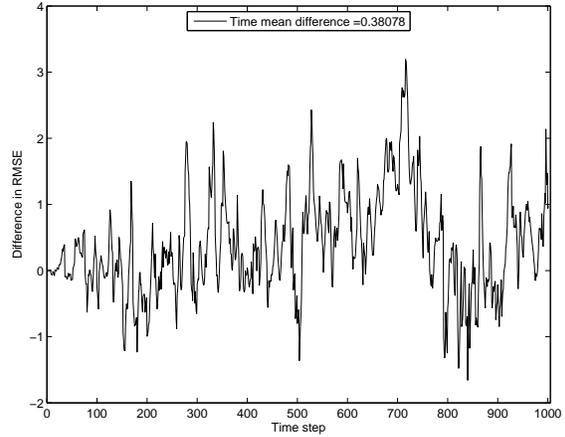}
     }
\caption{ \label{fig:rmse_plain_setting} Time series of the RMSEs in four different scenarios. For comparison, in Panels (b-d) we plot the differences of the RMSEs relative to those in Panel (a) (obtained by subtracting the RMSEs in (b-d) from those in (a)). In (b), the RMSEs overlap those in (a) up to around the first 130 integration steps. However, due to the chaotic nature of the ms-L96, tiny differences due to numerical precision are accumulated and amplified, and eventually become noticeable. In (c-d), due to the extra numerical errors in the DS-divided scenarios, the RMSEs are indistinguishable from those in (a) up to the first few integration steps only, and become noticeably different afterwards.}
\end{figure*}

\clearpage
\newcommand{\nscale}{0.4}
\newcommand{\xscale}{0.38}
\begin{figure*} 
\centering
	\subfigure[Boxplots (DS-joint,DA-divided)]{
	\label{fig:boxplot_sysSplit_estSplit_trajectory_Diff_DS-joint_DA-divided}	
	\includegraphics[scale = \nscale] {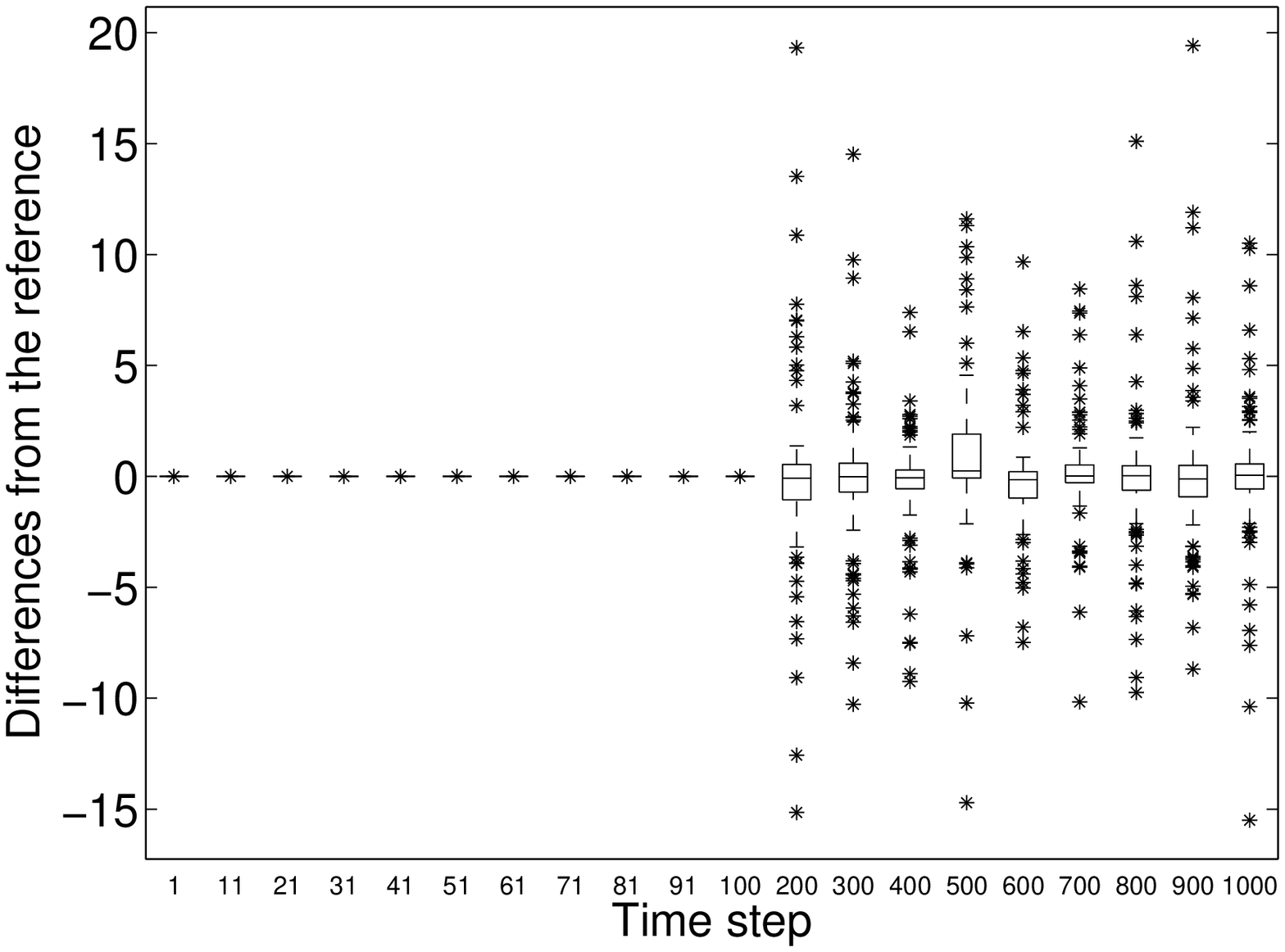}
     } 
	\subfigure[Histograms (DS-joint,DA-divided)]{
	\label{fig:histogram_DS-joint_DA-divided}	
	\includegraphics[scale = \xscale] {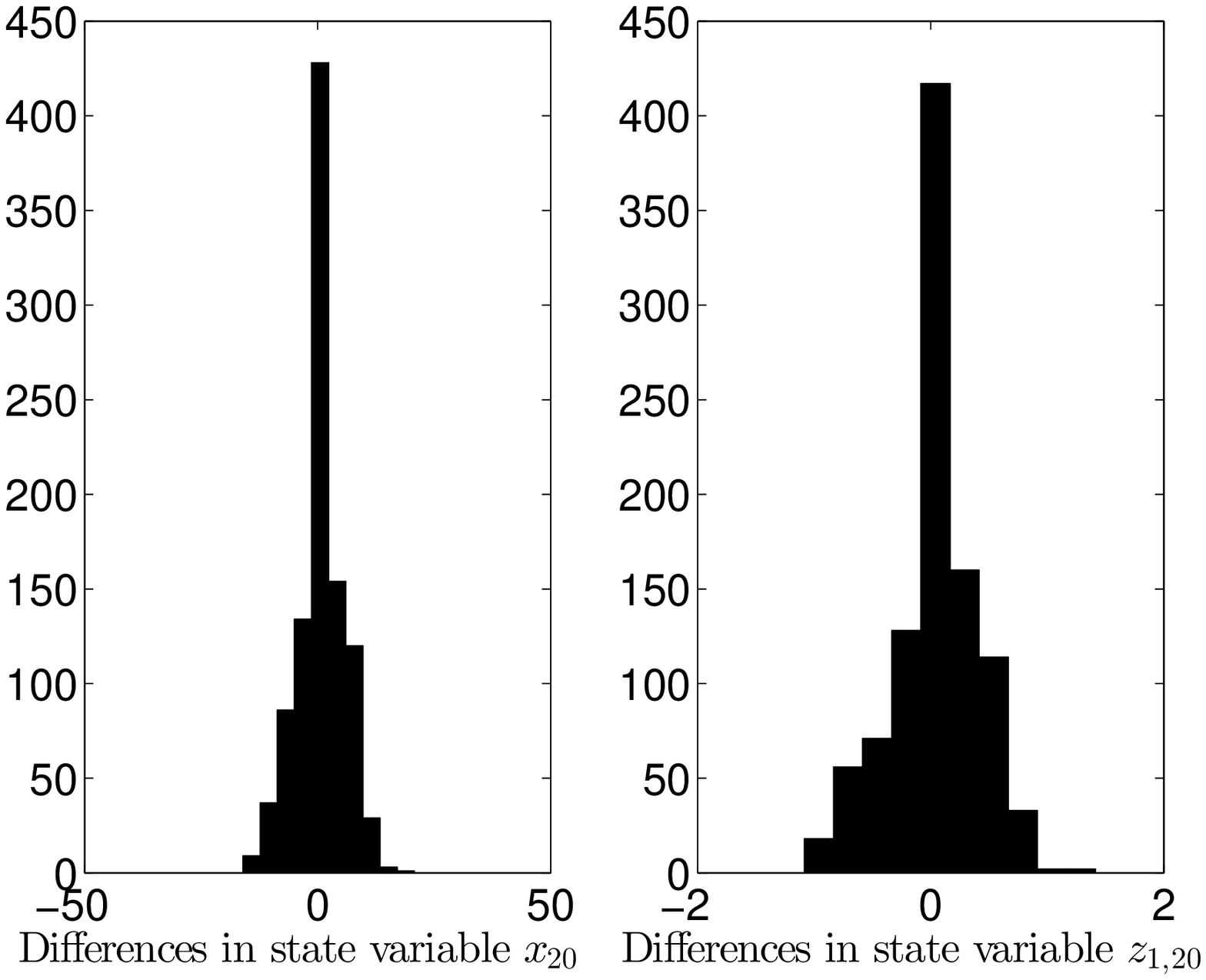}
     } \hspace{-0.0cm}     
    
	\subfigure[Boxplots (DS-divided,DA-joint)]{
	\label{fig:boxplot_sysSplit_estSplit_trajectory_Diff_DS-divided_DA-joint}	
	\includegraphics[scale = \nscale] {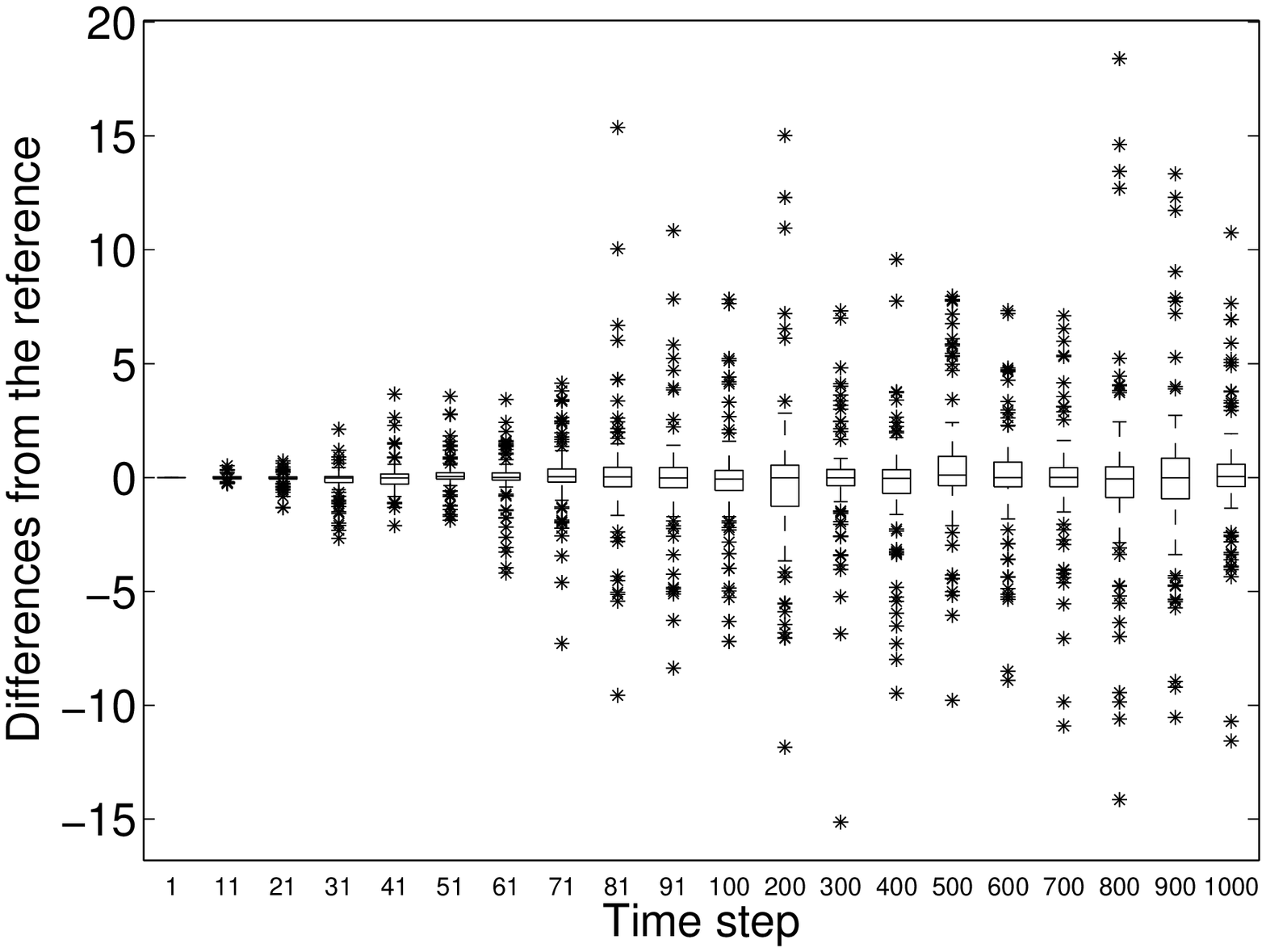}
     }
	\subfigure[Histograms (DS-divided,DA-joint)]{
	\label{fig:histogram_DS-divided_DA-joint}	
	\includegraphics[scale = \xscale] {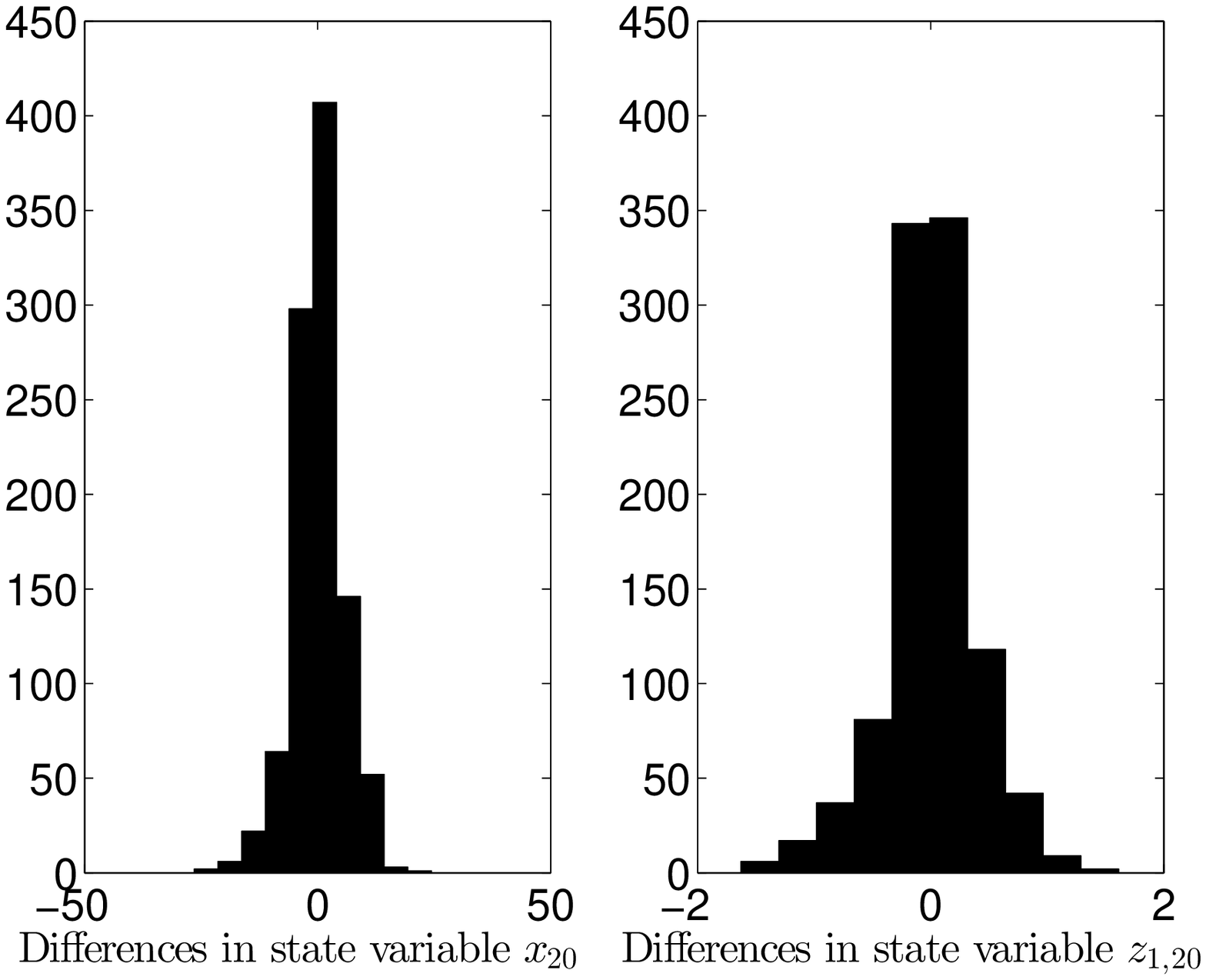}
     }    
     
	\subfigure[Boxplots (DS-divided,DA-divided)]{
	\label{fig:boxplot_sysSplit_estSplit_trajectory_Diff_DS-divided_DA-divided}	
	\includegraphics[scale = \nscale] {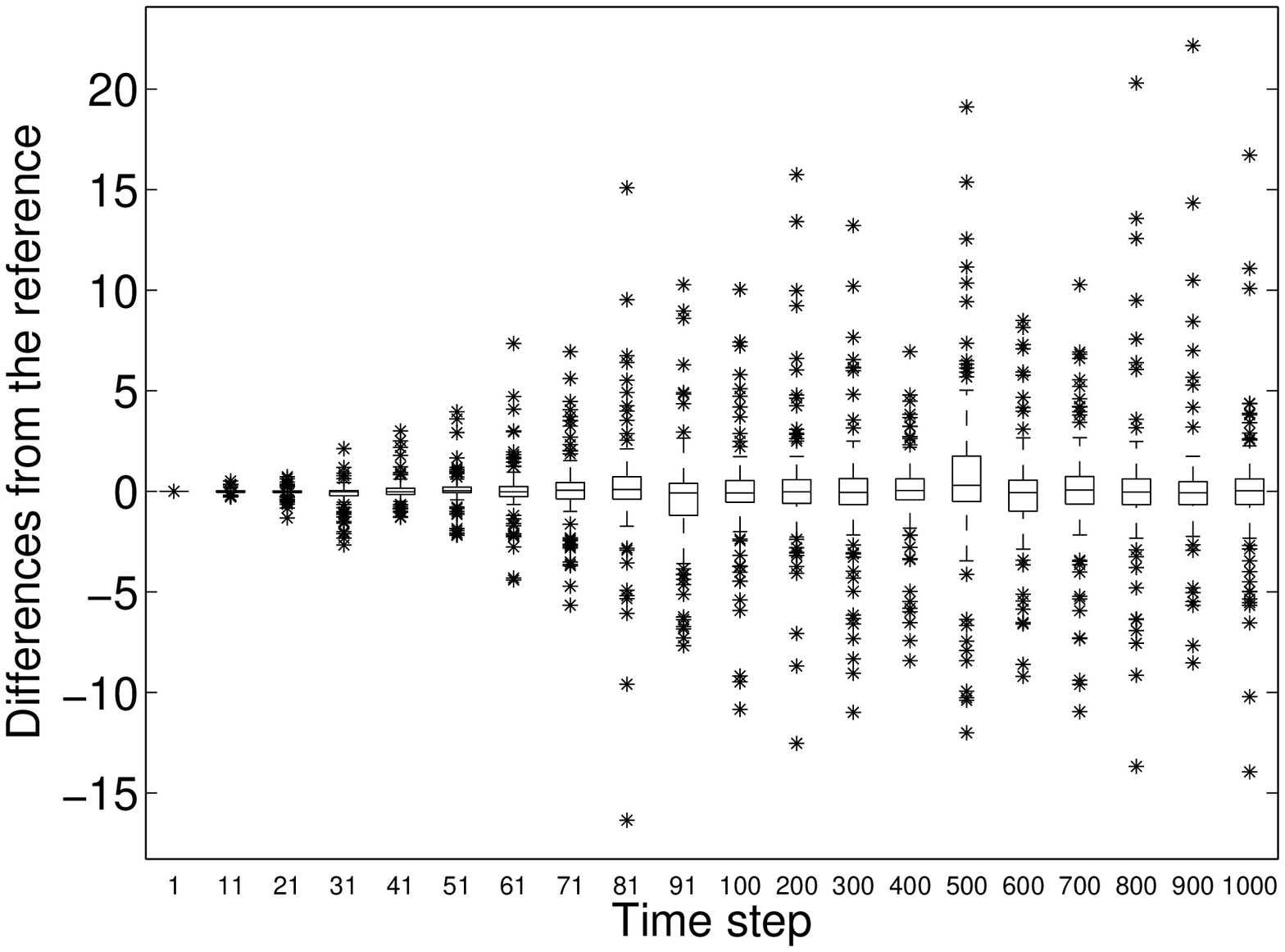}
     }
	\subfigure[Histograms (DS-divided,DA-divided)]{
	\label{fig:histogram_DS-divided_DA-divided}	
	\includegraphics[scale = \xscale] {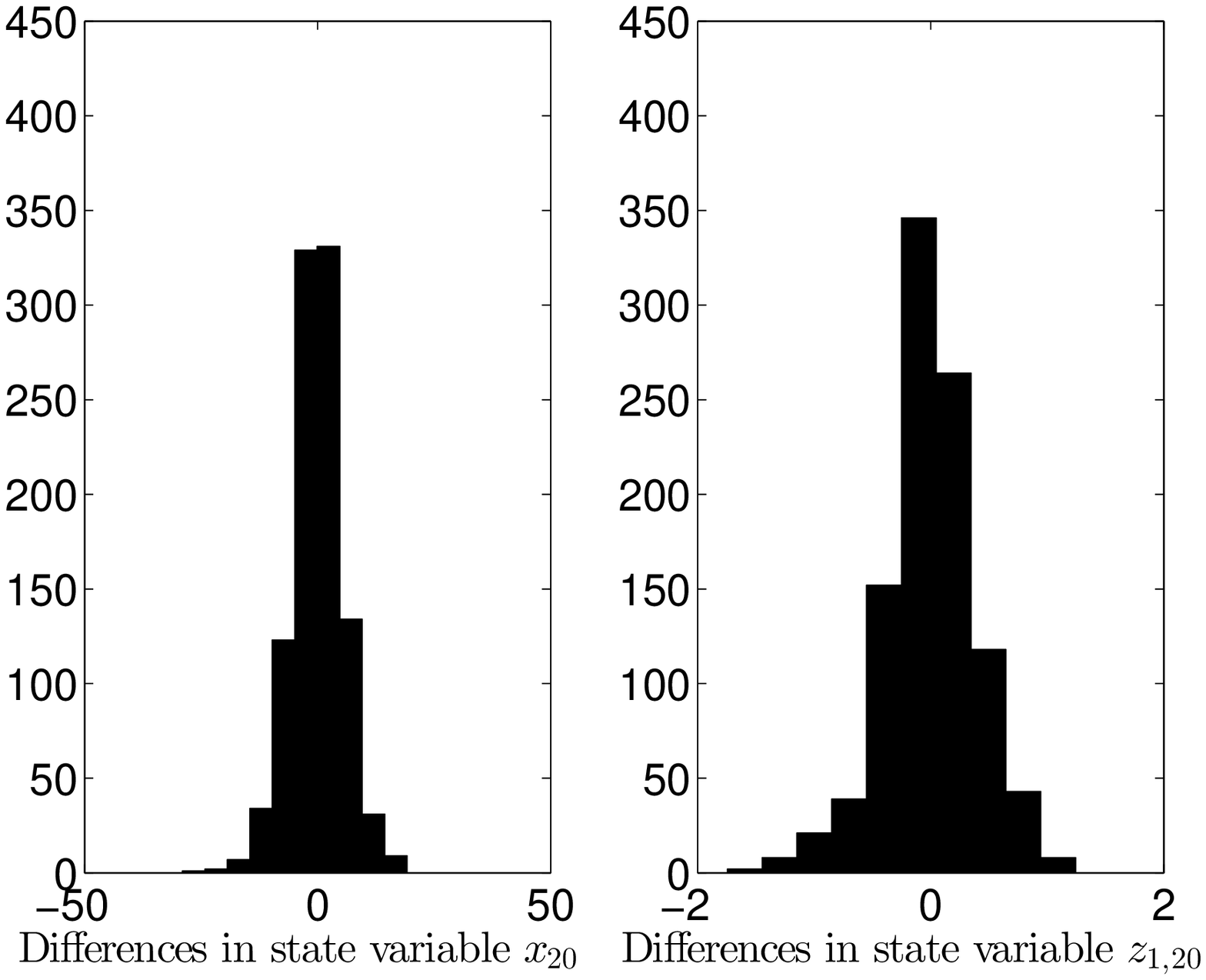}
     }    
     
\caption{ \label{fig:diff_box_plot} Boxplots (left column) and histograms (right column) for the characterization of the trajectory differences of the (DS-joint,DA-divided), (DS-divided,DA-joint) and (DS-divided,DA-divided) scenarios from the reference trajectory in the (DS-joint,DA-joint) scenario.}
\end{figure*}

\clearpage
\begin{figure*} 
\centering
	\subfigure[Scenario (DS-joint,DA-joint)]{
	\label{fig:rmse_noSysSplit_noEstSplit_varying_delta_and_lc}	
	\includegraphics[width = 0.45\textwidth] {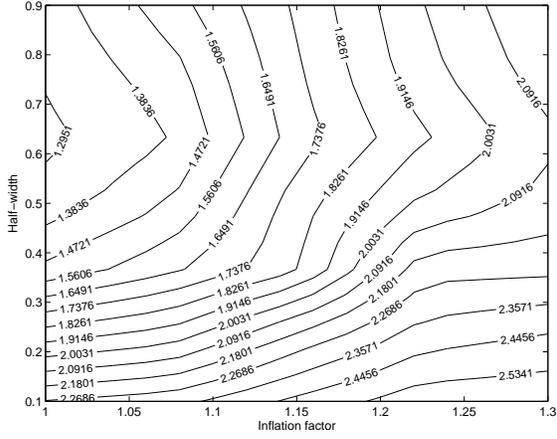}
     }
	\subfigure[Scenario (DS-joint,DA-divided)]{
	\label{fig:rmse_noSysSplit_estSplit_varying_delta_and_lc}	
	\includegraphics[width = 0.45\textwidth] {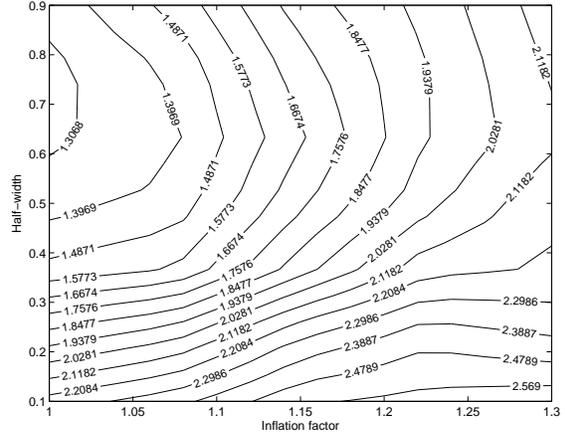}
     } \hspace{-0.0cm}
	\subfigure[Scenario (DS-divided,DA-joint)]{
	\label{fig:rmse_sysSplit_noEstSplit_varying_delta_and_lc}	
	\includegraphics[width = 0.45\textwidth] {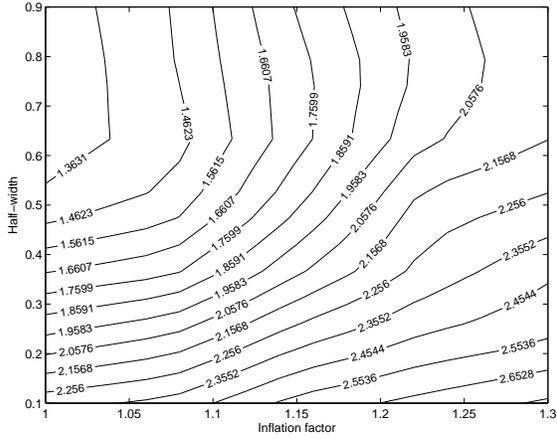}
     }
	\subfigure[Scenario (DS-divided,DA-divided)]{
	\label{fig:rmse_sysSplit_estSplit_varying_delta_and_lc}	
	\includegraphics[width = 0.45\textwidth] {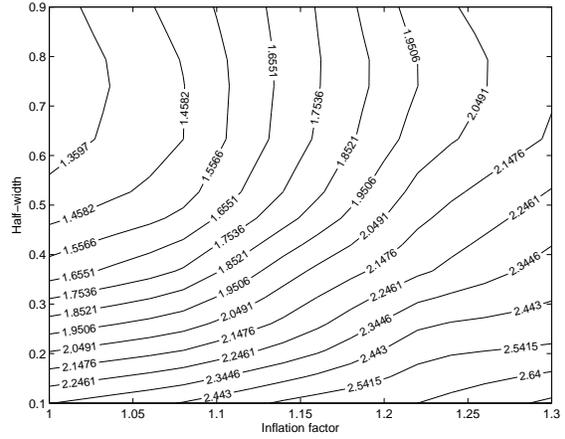}
     }
\caption{ \label{fig:rmse_lc_delta} Time mean RMSEs of four different scenarios with both covariance inflation and localization adopted. 
}
\end{figure*}

\clearpage
\begin{figure*} 
\centering
	\subfigure[Case $(n_f = 20,n_s =20)$]{
	\label{fig:L96_ensize_20_20}	
	\includegraphics[width = 0.45\textwidth] {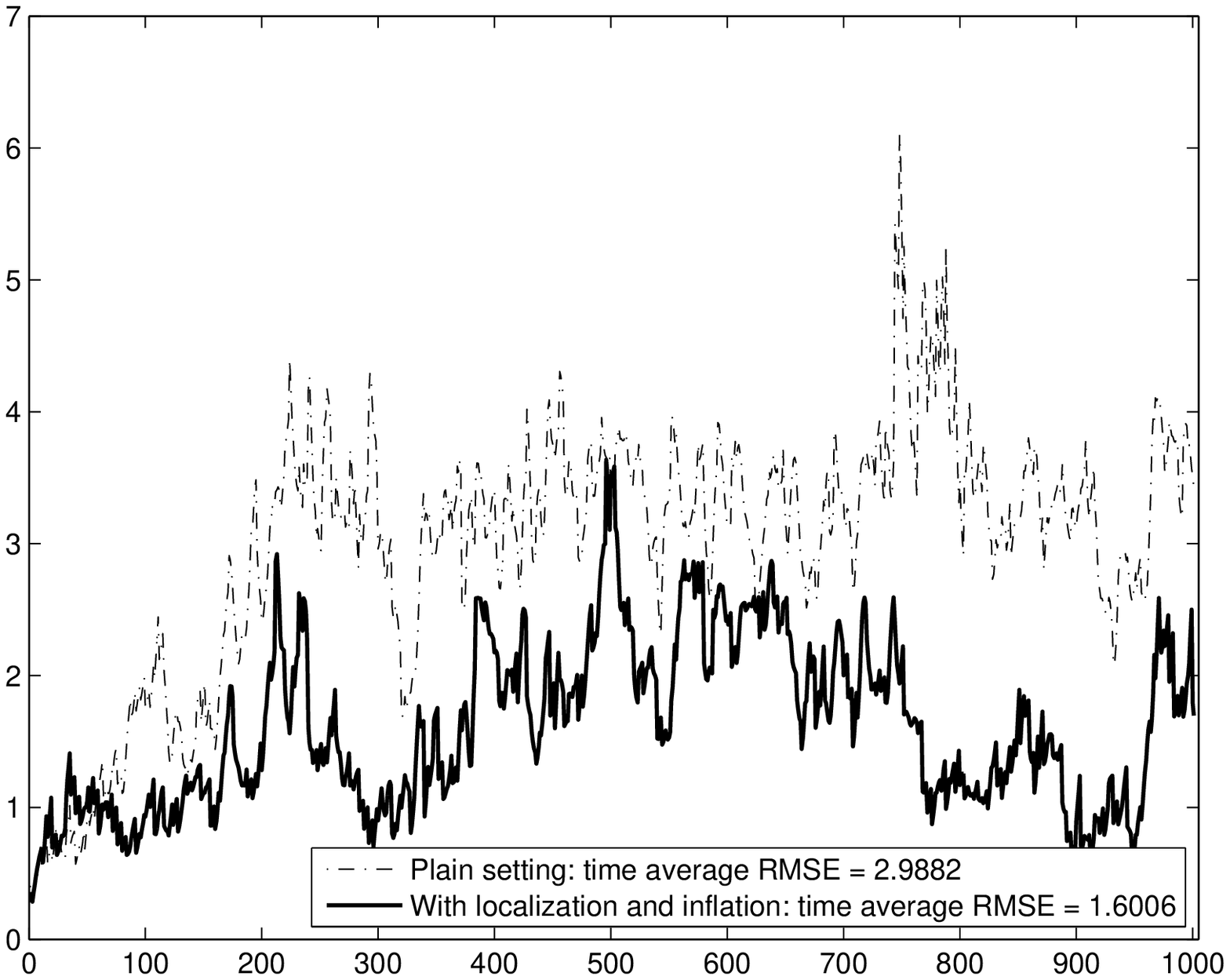}
     }
	\subfigure[Case $(n_f = 20,n_s =15)$]{
	\label{fig:L96_ensize_20_15}	
	\includegraphics[width = 0.45\textwidth] {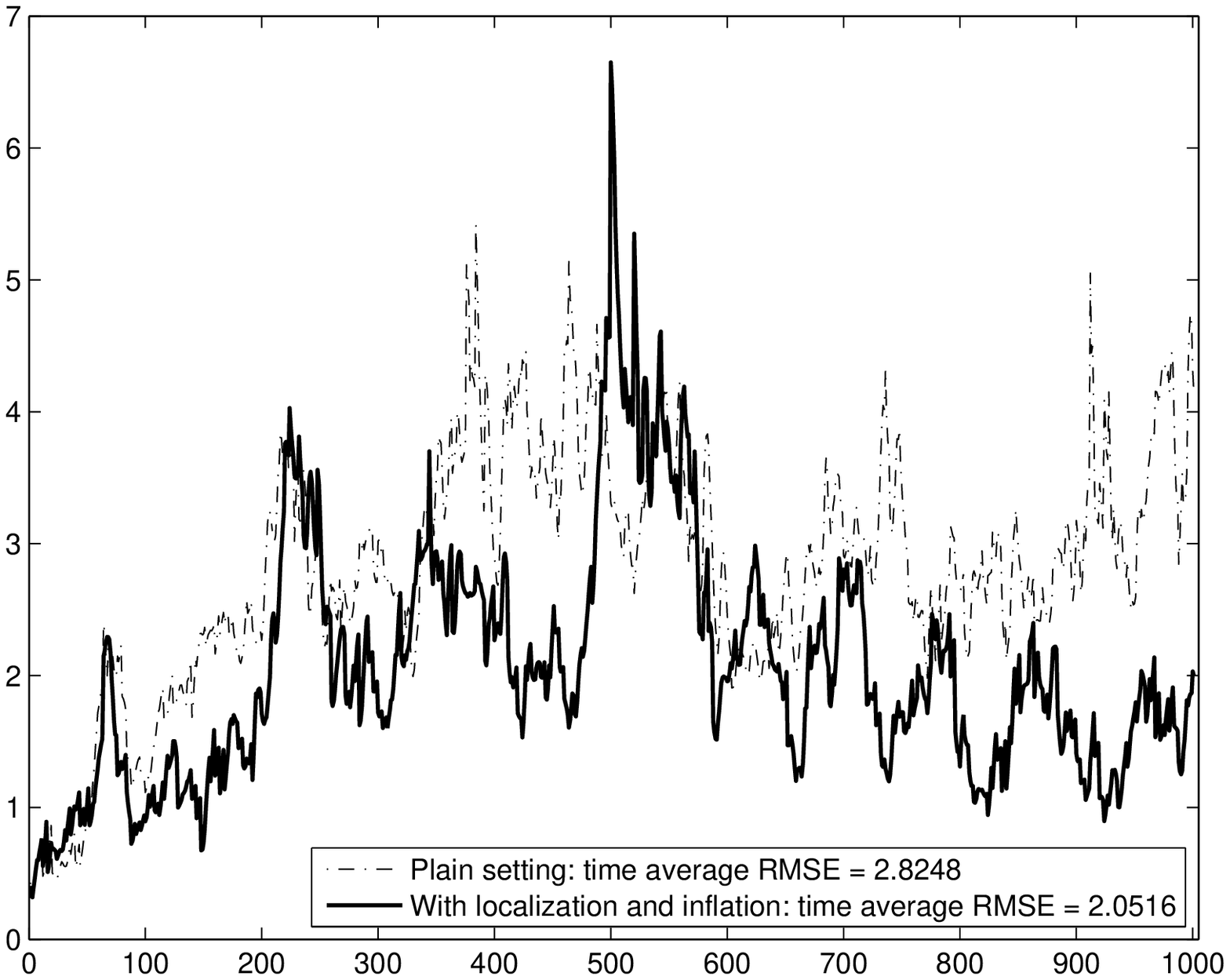}
     } \hspace{-0.0cm}
	\subfigure[Case $(n_f = 15,n_s =20)$]{
	\label{fig:L96_ensize_15_20}	
	\includegraphics[width = 0.45\textwidth] {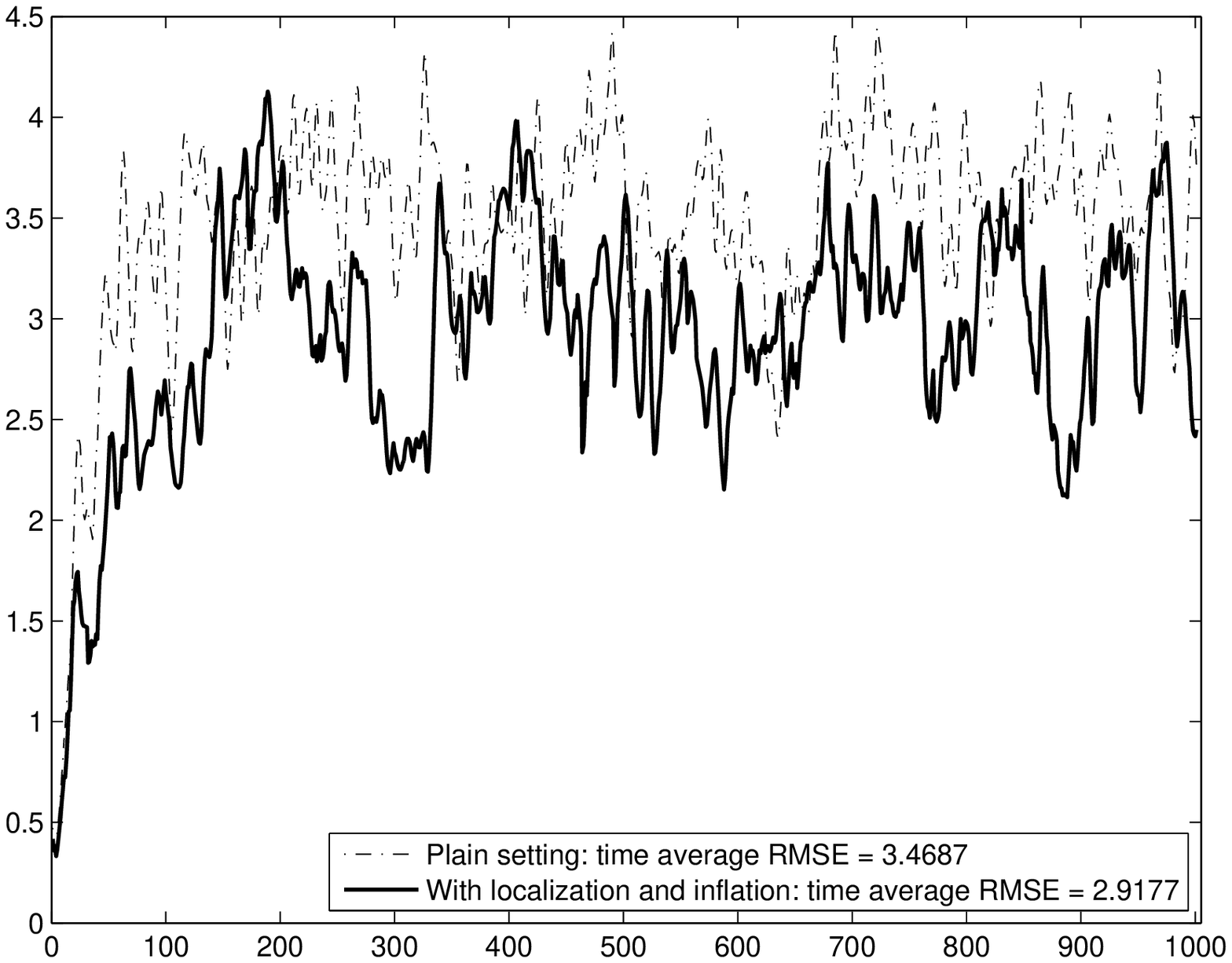}
     }
	\subfigure[Case $(n_f = 15,n_s =15)$]{
	\label{fig:L96_ensize_15_15}	
	\includegraphics[width = 0.45\textwidth] {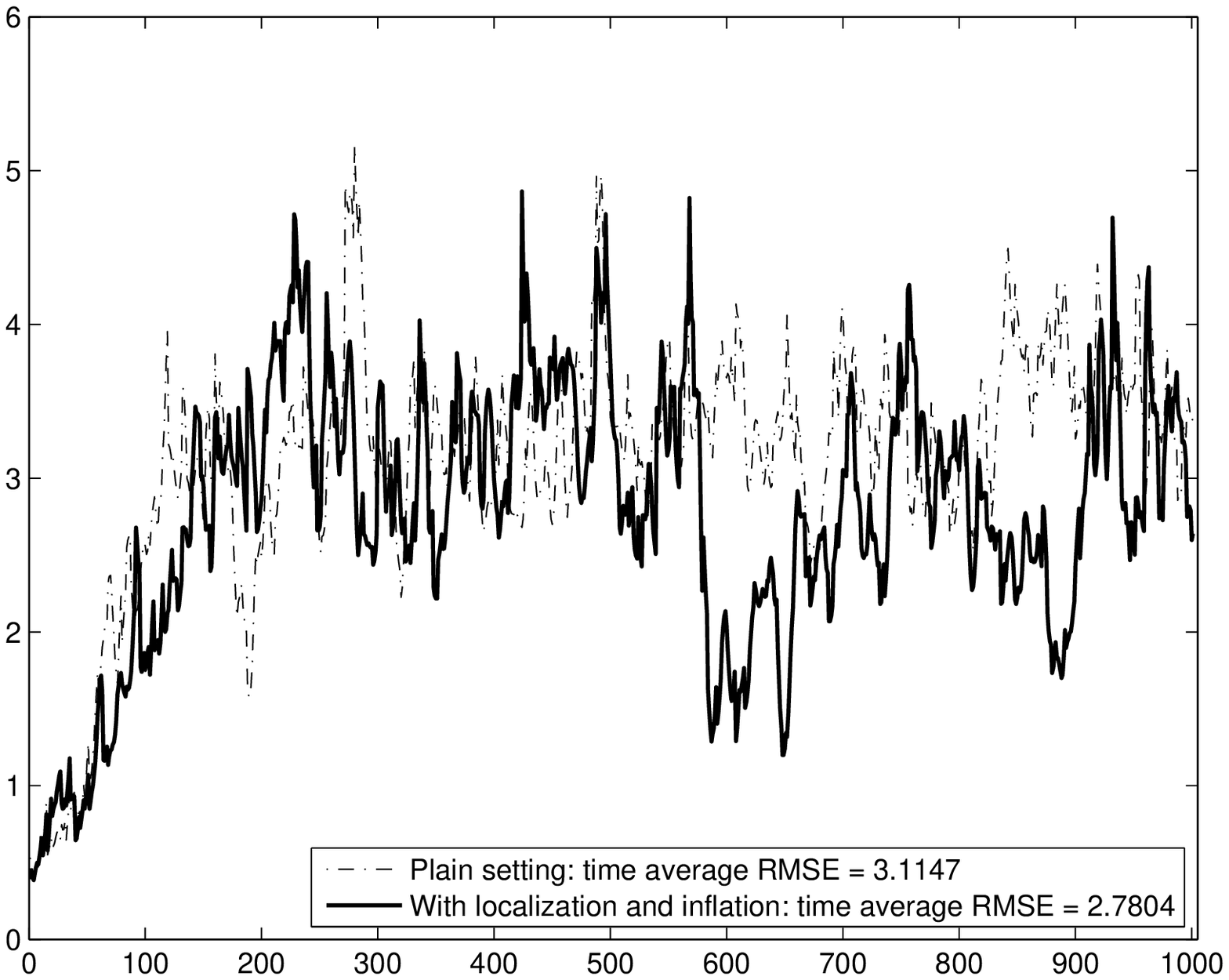}
     }
\caption{ \label{fig:rmse_ensize} Time series of the RMSEs of four cases with different ensemble sizes in the fast and slow modes.}
\end{figure*}

\clearpage
\begin{figure*} 
\centering
	\subfigure[``Climatological'' means]{
	\label{subfig:L96_lt_mean_values}	
	\includegraphics[width = 0.45\textwidth] {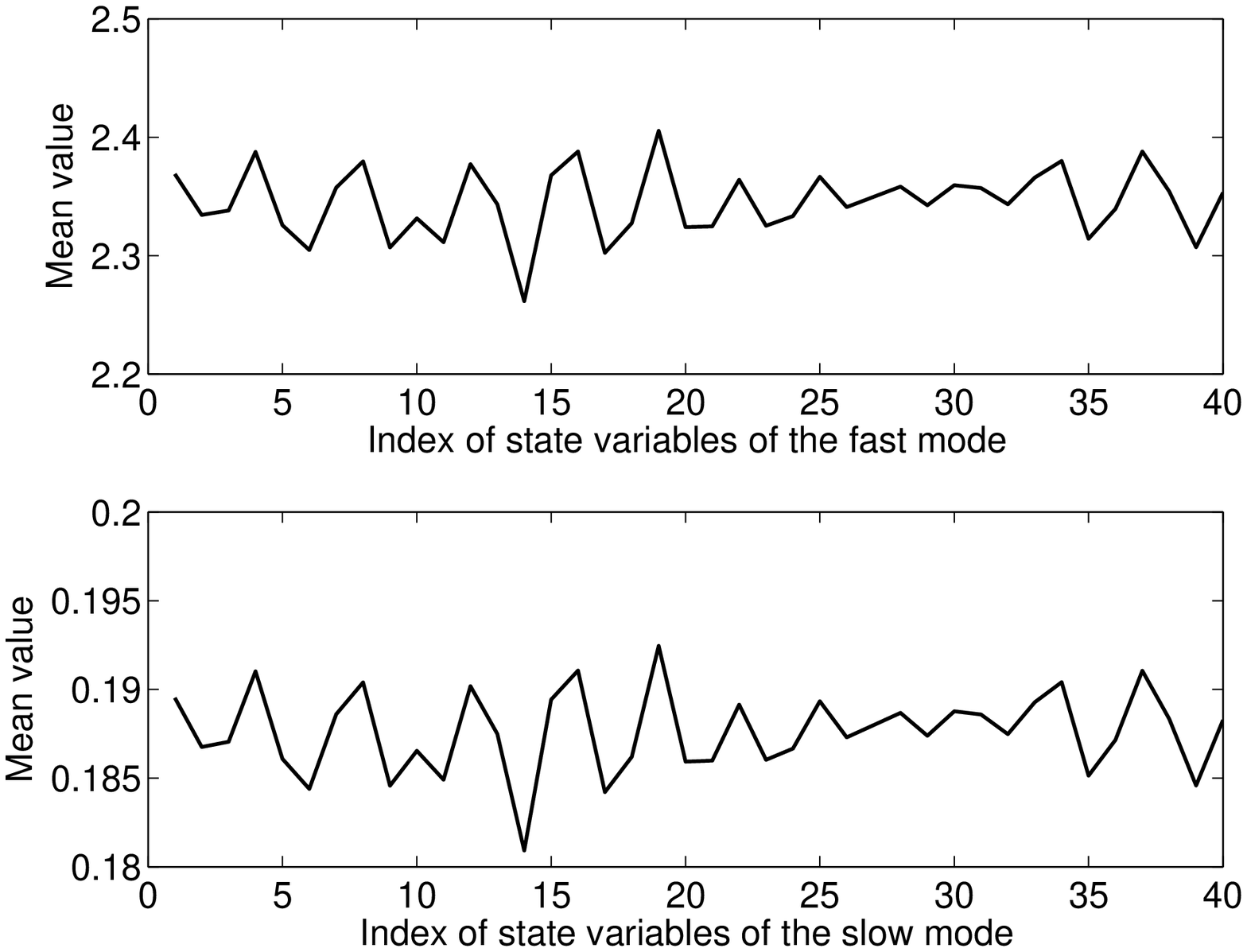}
     }
	\subfigure[Eigenvalues of the ``Climatological'' covariances]{
	\label{subfig:L96_Eigenvalues}	
	\includegraphics[width = 0.45\textwidth] {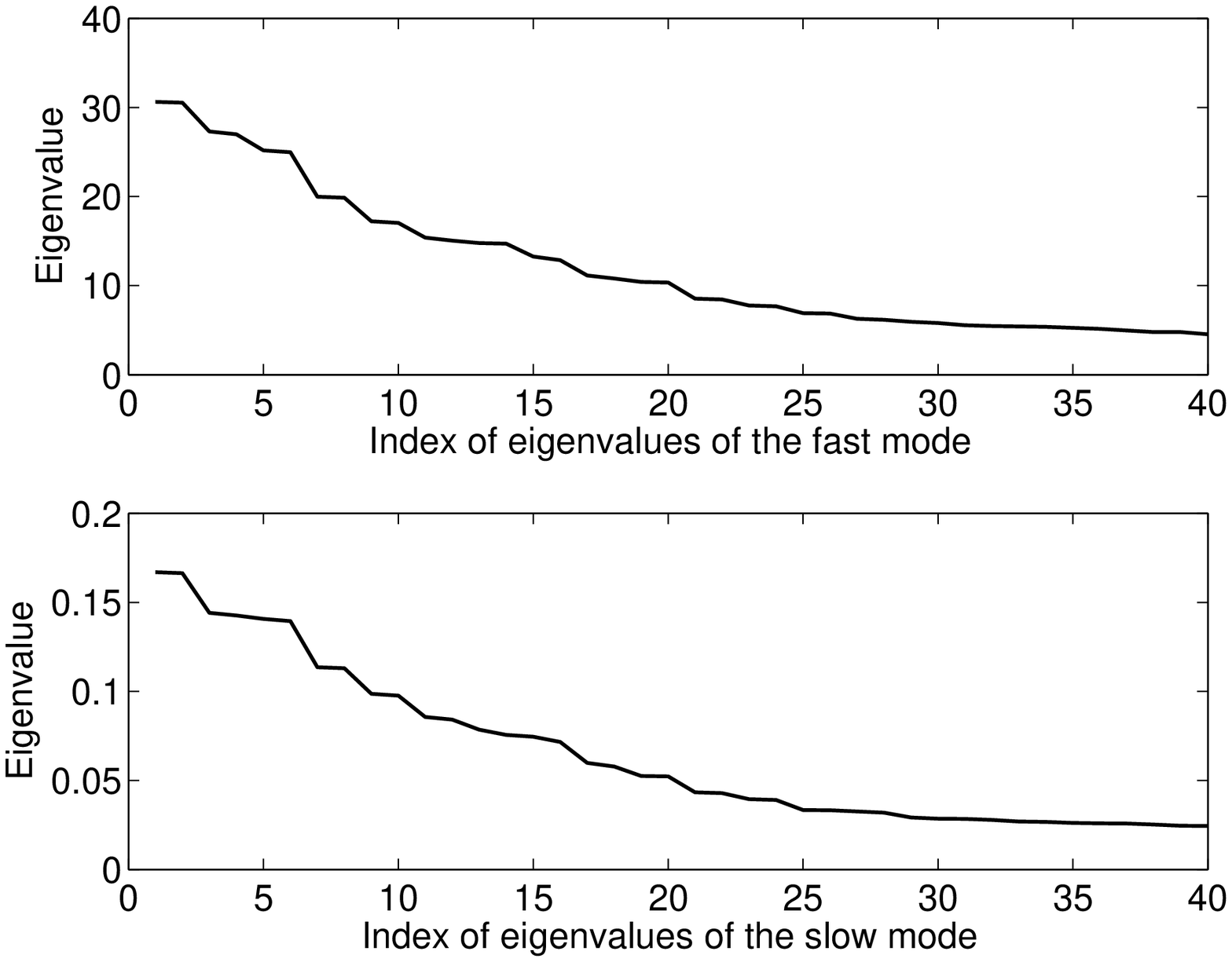}
     } \hspace{-0.0cm}
\caption{ \label{fig:lt_statistics} ``Climatological'' means and the eigenvalues of the ``climatological'' covariances of the fast and slow modes.}
\end{figure*}

\clearpage
\begin{figure*} 
\centering
	\subfigure[The plain setting]{
	\label{subfig:L96_plain_setting_EnOI}	
	\includegraphics[width = 0.8\textwidth] {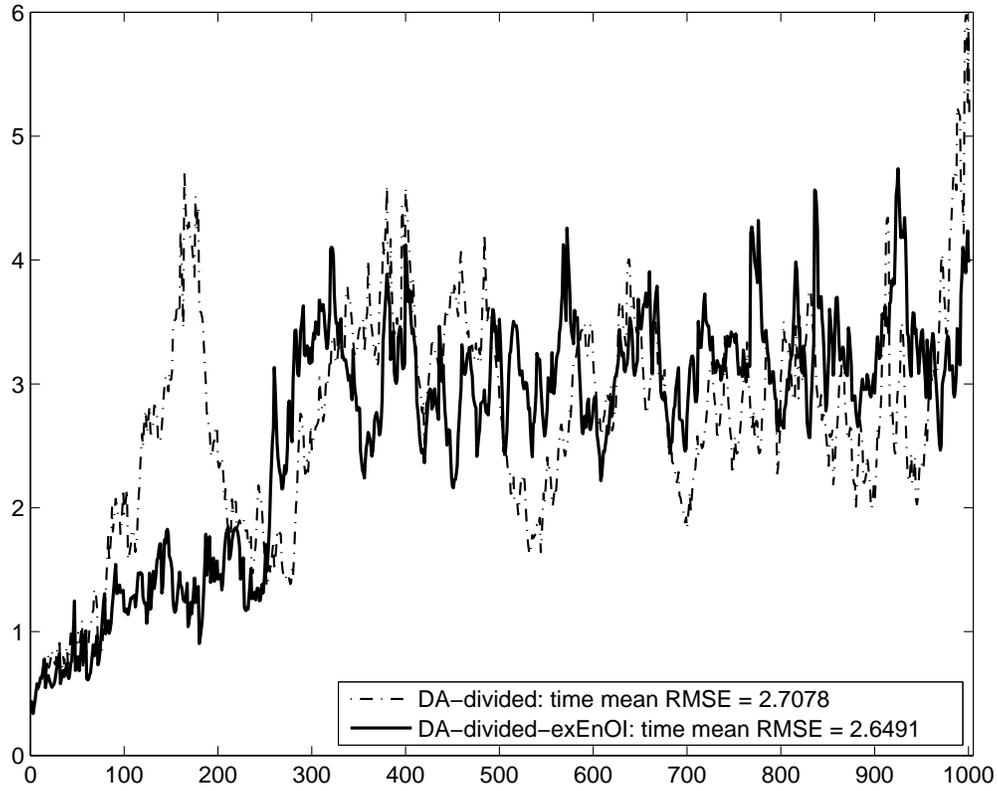}
     }
	\subfigure[With both covariance inflation and localizaton]{
	\label{subfig:L96_loc_inf_EnOI}	
	\includegraphics[width = 0.8\textwidth] {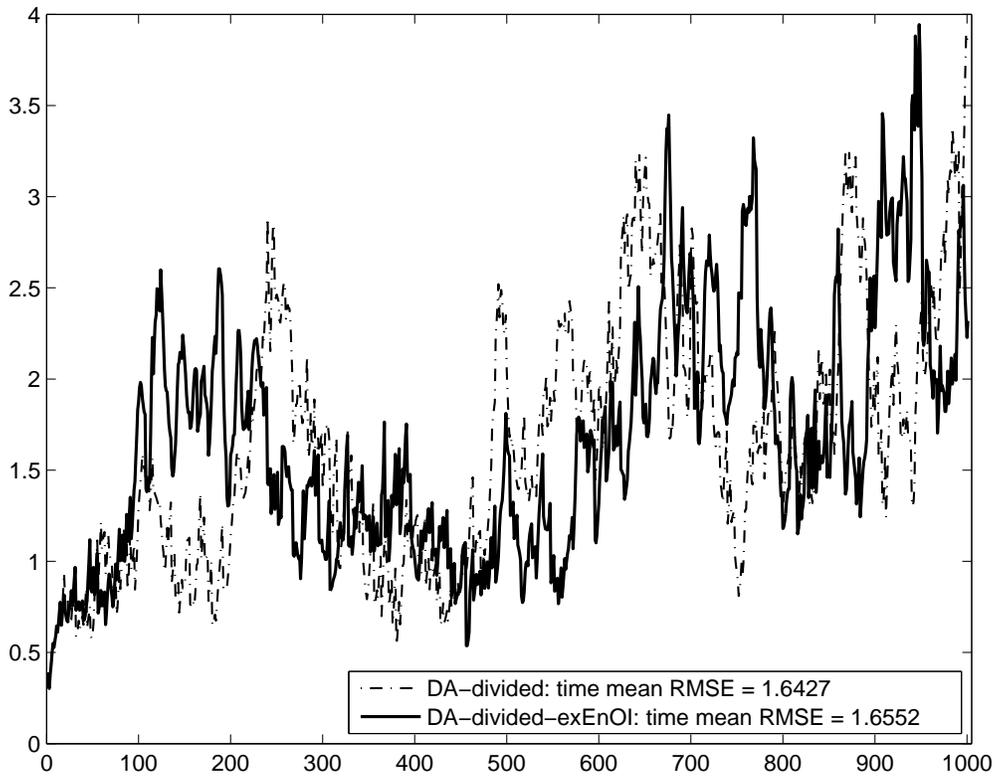}
     } \hspace{-0.0cm}
\caption{ \label{fig:EnOI} Time series of the RMSEs in cases of DA-divided and DA-divided-exEnOI. Panel (a) Neither covariance inflation nor localization is applied; Panel (b) both covariance inflation and localization are conducted.}
\end{figure*}

\end{document}